\documentclass{ar2e}

\newcommand{\bld}[1]{\mbox{\boldmath$#1$\unboldmath}}
\newcommand{\vesc}{{v_{\rm esc}}}
\newcommand{\lesssim}{\mathrel{\hbox{\rlap{\hbox{\lower4pt\hbox{$\sim$}}}\hbox{$<$}}}}
\newcommand{\gtrsim}{\mathrel{\hbox{\rlap{\hbox{\lower4pt\hbox{$\sim$}}}\hbox{$>$}}}}
\newcommand\apj{ {\it Ap. J.\ }}
\newcommand\apjl{ {\it Ap. J. Lett.\ }}

\newcommand\araa{ {\it Annu.\ Rev.\ Astron.\ Astrophys.\ }}

\newcommand\mnras{{\it MNRAS }}
\newcommand\planss{{\it  Planet. Space Sci. }}
\newcommand\pasp{{\it PASP }}
\newcommand\jgr{{\it J.\ Geophys.\ Res.\ }}
\newcommand\jrasc{ {\it J. R. Astron. Soc. Can.\ }}
\newcommand\nat{ {\it Nature\ }}
\newcommand\Icarus{{\it Icarus\ }}
\newcommand\Science{{\it Science\ }}
\newcommand\cm{{\, \rm cm}}
\newcommand\m{{\, \rm m}}
\newcommand\gm{{\, \rm g}}
\newcommand\km{{\, \rm km}} 
\newcommand\yr{{\, \rm y}}
\newcommand\Myr{{\, \rm Myr}}
\newcommand\au{{\, \rm AU}}
\newcommand\s{{\, \rm s}}

\begin{document}

\input epsf.tex 

\jname{Ann. Rev. Astron. Astrophys.}
\jyear{2004}
\jvol{42}
\ARinfo{1056-8700/97/0610-00}

\title{Planet Formation by Coagulation: A Focus on Uranus and Neptune}

\renewcommand{\baselinestretch}{1.00}

\markboth{Goldreich, Lithwick, and Sari}{Planet Formation}

\author{Peter Goldreich,$^{1,2}$ Yoram Lithwick,$^{3}$ and Re'em Sari$^{2}$
\affiliation{
$^1$Institute for Advanced Study, Princeton, New Jersey 08540\\
$^2$Theoretical Astrophysics, MC 130-33, Caltech, Pasadena, California 91125\\
$^3$Astronomy Department, University of California, Berkeley, California 94720 \\
email: pmg@ias.edu; yoram@astro.berkeley.edu; sari@tapir.caltech.edu
}}


\begin{abstract}

Planets form in the circumstellar disks of young stars.  We review the
basic physical processes by which solid bodies accrete each other and alter each
others' random velocities,  and we provide order-of-magnitude derivations
for the rates  of these processes. We discuss and exercise the two-groups approximation, a simple yet powerful technique for solving the
evolution equations for protoplanet growth. We describe orderly,
runaway, neutral, and oligarchic growth. We also delineate the conditions
under which each occurs. We refute a popular misconception by showing
that the outer planets formed quickly by accreting small bodies. Then
we address the final stages of planet formation.  Oligarchy ends when
the surface density of the oligarchs becomes comparable to that of the
small bodies. Dynamical friction is no longer able to balance viscous
stirring and the oligarchs' random velocities increase. In the inner-planet system, oligarchs collide and coalesce. In the outer-planet
system, some of the oligarchs are ejected. In both the inner- and outer-planet systems, this stage ends once the number of big bodies has been
reduced to the point that their mutual interactions no longer produce
large-scale chaos. Subsequently, dynamical friction by the residual
small bodies circularizes and flattens their orbits. The final stage
of planet formation involves the clean up of the residual small
bodies. Clean up has been poorly explored.

\end{abstract}

\maketitle

\small\normalsize
\parskip = 0.5em

\section{Introduction}

\footnotetext{Posted with permission, from the {\it Annual Review of Astronomy and Astrophysics}, Volume
42 \copyright  2004 by Annual Reviews, www.annualreviews.org}
The subject of planet formation is much too large for a short
review. Our coverage is selective. For a broader perspective, the
reader is encouraged to peruse other reviews  (e.g., \citen{K02}; \citen{Lis93}; 
Lissauer et al. 1995; 
Lissauer 2004; 
\citen{S72};  Wuchterl, Guillot \& Lissauer 2000). 
 Except for a description of
gravitational instabilities in cold disks in Section \ref{subsec:toomre}, we
bypass the crucial stage during which dust grains accumulate to form
planetesimals. Our story begins after planetesimals have appeared on
the scene.  Moreover, we focus on processes that occur in gas-free
environments. These were likely relevant during the later stages of
the growth of Uranus and Neptune. 

We devote the first half of the review to the basic physical processes
responsible for the evolution of the masses and velocity dispersions
of bodies in a protoplanetary disk.  Rather than derive precise
formulae governing the rates of these processes, we motivate
approximate expressions that capture the relevant physics and refer
the reader to more complete treatments in the literature
(\citen{BP90}; \citen{DT93}; Greenberg et al. 1991; Greenberg \& Lissauer 1990, 1992;
\citen{HN90}; Hornung, Pellat \& Barge 1985\nocite{HPB85}; \citen{IN89}; \citen{Oht99};
Ohtsuki, Stewart \& Ida 2002\nocite{Oht02}; Rafikov 2003a,b,d; \citen{SI00}
; Stewart \& Wetherill 1988).
Most of the basic physical processes are, by
now, well understood as the result of extensive analytical and
numerical investigations by many workers. The velocity dispersion in
the shear dominated regime is a notable exception, and here we
contribute something new.

The second half of the review is concerned with the growth of planets
starting from a disk of planetesimals.
Readers not interested in the derivations of equations describing mass and velocity evolution
can skip directly to the second half (beginning with Section \ref{sec:growthofplanets}).
Although many of our results
are general, we concentrate on the formation of the outer planets,
Uranus and Neptune. We make this choice for two reasons: The formation of Neptune and Uranus presents the most severe timescale problem. Yet it  might be the
simplest because it was probably completed in the absence of a
dynamically significant amount of gas.
Order-of-magnitude
estimates, particle-in-a-box simulations, and direct N-body
simulations have been used to address this problem. We briefly
review the weaknesses and strengths of each approach. Then we show how
the simplest of these, the two-groups approximation, can capture many of the results obtained in more
sophisticated treatments.
We show that the evolution of the mass spectrum can be either orderly, neutral, or runaway and
 describe oligarchy. We conclude with a discussion of how the Solar System evolved from its state at the end
of oligarchy to its present state.

 \section{The Hill Sphere}
\label{sec:thehillsphere}
For clarity, we consider interactions between two sets of bodies: big
ones (protoplanets or embryos)\footnote{We use these terms
interchangeably.} with mass $M$ and small ones (planetesimals) with
mass $m$.  See Table \ref{table:symbols} for a list of symbols.  
Small bodies have random velocity $u$, i.e., eccentricity 
$\sim u/\Omega a$, and big bodies have random velocity $v$.
We assume, unless explicitly stated otherwise, that 
inclinations are comparable to eccentricities, in which case the small and big bodies have
vertical scale heights
 $u/\Omega$ and $v/\Omega$,
respectively.  We also assume that $v<u$, unless stated
otherwise (e.g., in Section \ref{s:v>u}).


\subsection{Hill Radius, $R_H$, and Hill Velocity, $v_H$}
\label{subsec:hill}

Planetary accretion calls into play a number of special solutions to
the restricted 
three-body problem.  An important system consists of the Sun, a big
body, and a small body. The Sun has mass and radius $M_\odot$ and
$R_\odot$, the big body has mass and radius $M$ and $R$, and the small
body is here treated as a massless test particle.
If the small body is close enough to the big body, then the Sun's
tidal gravitational field is negligible relative to that of the big
body. Conversely, if the small body is far enough away, then the big
body's 
gravitational
field is negligible relative to that of the Sun. The Hill
radius, $R_H$, is the characteristic distance from the big body that
distinguishes between these two behaviors.  At a distance $R_H$, the
orbital frequency around the big body is comparable to the orbital
frequency of the big body around the Sun, i.e.,
$(GM/R_H^3)^{1/2}\sim(GM_\odot/a^3)^{1/2}\sim \Omega$,
where $a$ is the distance to the Sun.  Therefore, $R_H\sim
a(M/M_\odot)^{1/3}$, or
\begin{equation}
R_H\sim R/\alpha \ ,
\label{eq:hillradius}
\end{equation}
where  
\begin{equation}
\alpha\equiv \left(\rho_\odot\over\rho\right)^{1/3}{R_\odot\over a}\,\, .
\label{eq:defalpha}
\end{equation}
Here $\rho_\odot$ and $\rho$ are  
the mean densities of the Sun and of a solid 
body, respectively.
Because $\rho\approx \rho_\odot\approx 1\gm\cm^{-3}$, $\alpha$ is
the approximate angular diameter of the Sun as seen from distance
$a$. From Earth $\alpha\sim 10^{-2}$, and from the
Kuiper Belt $\alpha\sim 10^{-4}$. 
A parameter similar to our $\alpha$ has been
used by  Dones \& Tremaine (1993), Greenzweig \& Lissauer (1990), and Rafikov (2003c). 

The Hill velocity $v_H$ is a characteristic
velocity associated with $R_H$.  It is the orbital velocity around the big body at a distance $R_H$, i.e.,
$v_H\sim (GM/R_H)^{1/2}$ or
\begin{equation}
v_H\sim \vesc \alpha^{1/2}\sim \Omega R_H \ .
\label{eq:hillvelocity}
\end{equation}

For $u>v_H$, close encounters of small bodies with big ones are well
approximated by two-body dynamics, but for $u<v_H$, the tidal gravity of
the Sun must be taken into account. The
former regime is
referred to as
 dispersion dominated and the latter is 
referred to as shear
dominated.

Small bodies that are initially on circular orbits around the Sun
(i.e., $u=0$) and pass near a big body show three types of behavior
(see figure 1 in Petit \& Henon 1986):
For impact
parameters from the big body less than approximately $R_H$, small bodies are reflected
in the frame of the big body upon approach and travel on horseshoe
orbits.  For impact parameters greater than a few times $R_H$, they
suffer small deflections while passing the big body.  For intermediate
impact parameters, they enter
the big body's Hill sphere, i.e., the sphere of radius $\sim R_H$
around the big body.  Inside the Hill sphere, small bodies follow complex
trajectories. Most do not suffer physical collisions with the big body
and exit the Hill sphere in an
arbitrary direction with random velocities of order $v_H$.

\subsection{Hill Entry Rate When $u<v_H$}

When $u<v_H$, the number of small bodies that enter into a given big body's Hill
sphere per unit time (the Hill entry rate) is an
essential coefficient for estimating the rates of 
evolution discussed in Sections \ref{sec:mass} and  \ref{sec:velevo}. 
As long as $u<v_H$, the scale height of the disk of
small bodies ($u/\Omega$) is smaller than $R_H$.  So to
calculate the rate at which small bodies enter the Hill sphere, we
approximate their disk as being infinitely thin.  Their entry rate is
equal to
their number per unit area ($\sigma/m$, where $\sigma$ is their surface 
mass density)
multiplied by $v_H R_H$, because the Keplerian shear velocity with which
small bodies approach the Hill sphere is $\sim v_H$ and the range of
impact parameters within which small bodies enter the Hill sphere is
$\sim R_H$.  Combining the above terms, we have

\begin{equation}
{\rm Hill \ entry \ rate}\sim {\sigma\over m}\Omega R_H^2 \ \ .
\label{eq:hillentryrate}
\end{equation}

\section{Collision Rates}
\label{sec:mass}

Here we calculate the collision rate suffered by a single big body embedded
in a disk of small bodies as the entire swarm orbits the Sun.  In Section
\ref{sec:massgrowthrate}, we convert the collision rate to a mass
growth rate, $dM/dt$. Because $v<u$, the orbit of the big body around the Sun
can be treated as though it were circular
($v=0$).  
There are four regimes of interest, depending on the value of $u$ (see Figure 1).


\subsection{Super-Escape: $u>\vesc$}
For $u>\vesc$, where $\vesc$ is the escape speed from the surface of
the big body, gravitational focusing is negligible, and the
collisional cross section is $R^2$, where $R$ is the radius of the big
body.  We denote the surface mass density of small bodies by 
$\sigma$, so their volumetric number density is
$\sigma\Omega/(mu)$.  Thus,
\begin{eqnarray}
{\rm collision\ rate} \sim
{\sigma\Omega\over m u}R^2 u =  {\sigma\Omega\over m}R^2
\ \ . \ \ u>\vesc.
\label{eq:cr1}
\end{eqnarray}
The collision rate depends on the surface number density of
small bodies because the $u$ dependence of their flux is canceled by
that from their scale height.

\subsection{Sub-Escape and Super-Hill: $\vesc>u>v_H$}

For $u<\vesc$, gravitational focusing enhances the collisional
cross section.  The impact parameter $b_{\rm graze}$ with which a
small body must approach the big one just to graze its
surface is determined as follows: At
approach the small body's angular momentum per unit mass around the
big body is $ub_{\rm graze}$; at contact, it is $\vesc R$. Thus
\begin{equation}
b_{\rm graze}\sim R{\vesc \over u} \ , 
\label{eq:graze}
\end{equation}
and
the collisional cross section is 
$b_{\rm graze}^2\sim R^2(\vesc/u)^2$ \,.
Thus gravitational focusing enhances the collision rate
given in Equation \ref{eq:cr1} by $(\vesc/u)^2$:
\begin{eqnarray}
{\rm collision\ rate} \sim {\sigma\Omega\over m}R^2
\left({\vesc\over u}\right)^2 \ \ ; \ \ \vesc>u>v_H \ .
\label{eq:cr2}
\end{eqnarray}

\subsection{Sub-Hill and Not Very Thin Disk: $v_H>u>\alpha^{1/2}v_H$}

For $u<v_H$, the gravitational field of the Sun affects the collision
rate, and the Keplerian shear affects both entry and exit from the
Hill sphere.  Greenberg et al. (1991) were the first to derive
analytical expressions for the collision rate in this
regime. 
Although based on their findings, our formulae differ slightly: 
For reasons that we feel are not well-justified, they  used
the Tisserand radius, whereas we use the Hill radius. 
 The former is smaller than the latter by $\sim(M/M_\odot)^{1/15}$.

The collision rate is the product of two terms: 
\begin{equation}
{\rm collision \ rate} \sim ({\rm Hill \ entry \ rate}) \cdot {\rm P},
\label{eq:collisionratesubhill1}
\end{equation}
where the Hill entry rate (Equation 4) is the rate at which small bodies enter the
big body's Hill sphere and P is the probability that, once inside the
Hill sphere, the small body impacts the big body.

\begin{figure}
\centerline{\epsfxsize=5.5in\epsfbox{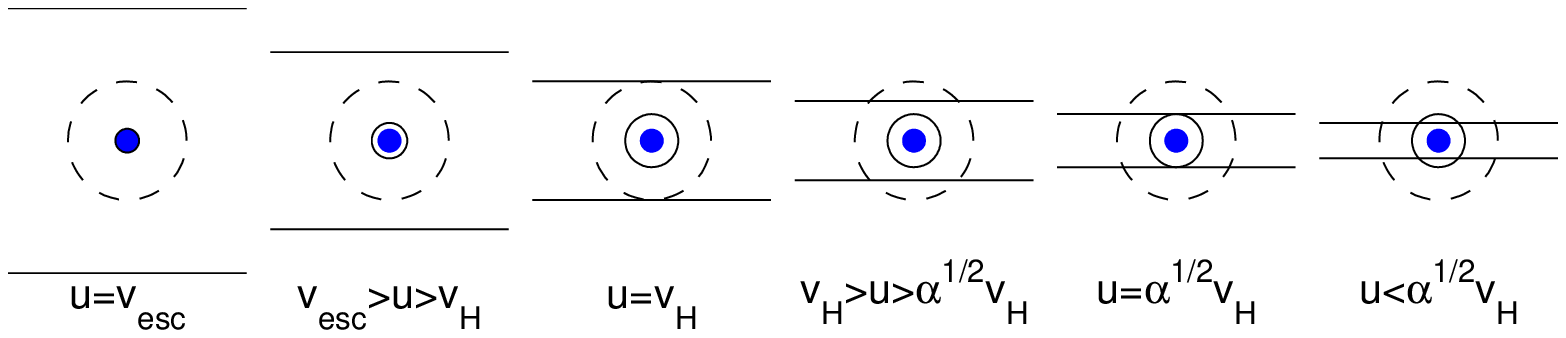}}
\caption{Geometry of disk scale height ({\it solid horizontal line}), body
size ({\it filled circle}), its Hill sphere ({\it dashed circle}), and its
effective size for accretion ({\it solid circle}) 
\label{fig:rates}}
\end{figure}

We estimate $P$ as follows: A small body that enters the Hill sphere
has its random velocity boosted from $u$ to $v_H$.\footnote{
The
increase is primarily in the horizontal components of the random
velocity.  At most, the vertical component is doubled.}  Once inside
the Hill sphere, if the small body's impact parameter relative to the big body is less
than $b_{\rm graze}\sim R\vesc/v_H$ (see Equation \ref{eq:graze}), a
collision will occur.  Based on Equations \ref{eq:hillradius} and
\ref{eq:hillvelocity}, $b_{\rm graze}\sim \alpha^{1/2}R_H$, so
$b_{\rm graze}<R_H$. We now consider the case
in which $b_{\rm graze}$ is also smaller than the scale height of small
bodies, $u/\Omega$, i.e.,
\begin{equation}
u> \alpha^{1/2}v_H \ .
\end{equation}
Although we can consider the disk of small bodies to be infinitely
thin when calculating the Hill entry rate, we must account for its
finite thickness when calculating the collision probability $P$.
Because the small bodies' trajectories within the Hill sphere are random,
we estimate $P$ as the ratio of the collisional cross section
($b_{\rm graze}^2$) to the cross section of the portion of the Hill
sphere that lies within the disk of small bodies ($R_H
u/\Omega$)
(see
Figure \ref{fig:rates}). We obtain
\begin{equation} 
P\sim {R_H^2 \alpha\over R_H u/\Omega}
\sim \alpha {v_H\over u} 
\ \ , \ \ v_H>u>\alpha^{1/2}v_H
\ .
\label{eq:P}
\end{equation}
Inserting Equations \ref{eq:hillentryrate} and \ref{eq:P} into
Equation \ref{eq:collisionratesubhill1}, we find
\begin{equation}
{\rm collision\ rate}  \sim
{\sigma\Omega\over m}R^2\alpha^{-1}{v_H\over u}
\ \ ; \ \ v_H>u>\alpha^{1/2}v_H \ .
\label{eq:cr3}
\end{equation}

\subsection{Very Thin Disk: $u<\alpha^{1/2} v_H$}

For $u<\alpha^{1/2}v_H$, the calculation proceeds as described above, except that the disk of small bodies is
considered infinitely thin in the estimation of $P$ (Greenberg et al. 1991, Dones \& Tremaine 1993).  Thus $P$ is
the ratio of the linear range of impact parameters leading to
collision ($b_{\rm graze}\sim R_H\alpha^{1/2}$) to $R_H$; i.e.,
\begin{equation}
P\sim \alpha^{1/2} \ \ , \ \ u<\alpha^{1/2} v_H \ .
\end{equation}
Inserting this into Equation \ref{eq:collisionratesubhill1} and
making use of Equation \ref{eq:hillentryrate} yields
\begin{equation}
{\rm collision\ rate}  \sim 
{\sigma\Omega\over m}R^2
\alpha^{-3/2}
\ \ ; \ \ u<\alpha^{1/2}v_H \ .
\label{eq:cr4}
\end{equation}

\section{Velocity Evolution}   
\label{sec:velevo}

Random velocities of big and small bodies, $v$ and $u$, evolve through
three processes: cooling by dynamical friction, heating by dynamical
friction, and viscous stirring. Terms that promote equipartition of
random kinetic energies (dynamical friction) may be separated from those that, in the
absence of dissipation, increase random kinetic energies (viscous stirring). In
astrophysics, dynamical friction conventionally refers to the
cooling term only. To avoid confusion, we always use the modifiers cooling and heating to refer to each effect separately.  It is
the combination of cooling and heating by dynamical friction that
promotes equipartition.

In addition to our usual assumptions, that $u>v$ and $m<M$, we also
require that $mu<Mv$.  This ensures that a big body will collide with
many small bodies before its velocity changes significantly;
otherwise, a single collision would suffice.  Although it is simple to
consider $mu>Mv$ as well, we do not because it is less
applicable to protoplanetary disks.

In this section, we quantify the rates of cooling and heating by
dynamical friction between groups of bodies with different masses. We
do the same for heating by viscous stirring. Unlike dynamical
friction, viscous stirring also increases the total energy within a
group of homogeneous bodies. We evaluate this in
Section \ref{sec:selfvs}.  In the following, separate
subsections are devoted to the different velocity regimes: $u>\vesc$,
$\vesc>u>v_H$, $u<v_H$. The most important results are gathered in
Section \ref{sec:massgrowthrate}.

\subsection{Super-Escape: $u>\vesc$ and Elastic Collisions}
\label{subsec:velubig}

For $u>\vesc$, gravitational focusing is negligible, and $v$ and $u$
change only when bodies collide. We assume elastic collisions and
neglect fragmentation because our aim is to develop intuition applicable
to collisionless gravitational scatterings (described below). This case is simple, yet it illustrates how
energy is transferred between big and small bodies.

\subsubsection{$dv/dt$ due to Dynamical Friction Cooling and Heating} 
\label{subsubsec:velubigdfv}

A big body that suffers a head-on collision with a small body
loses $\sim m(u+v)$ of its momentum.  Because the head-on collision
rate is $\sim n_s R^2(u+v)$, where $n_s$ is the
number density of small bodies, the big body loses momentum at the
rate
\begin{equation}
M{dv\over dt}\Big\vert_{\rm head-on} \sim -n_s R^2m (u+v)^2 
\end{equation}
as a result of such collisions. Conversely, it gains momentum at the rate
\begin{equation}
M{dv\over dt}\Big\vert_{\rm tail-on} \sim +n_s R^2m (u-v)^2 \ 
\end{equation}
owing to tail-on collisions.
Combining these two expressions, we find
\begin{equation}
{1\over v}{dv\over dt}\Big{\vert}_{\rm df\ cool}
\sim
-n_suR^2{m\over M} \ 
\label{eq:simplecool}
\end{equation}
to lowest order in $v$.  Therefore the big body slows down in the time
that it collides with a mass $M$ of small bodies.  

If the big body is  sufficiently slow, small bodies tend to speed it
up.  Each scattering with a small body transfers some momentum to the
big body, with roughly fixed amplitude ($\sim mu$) and random
orientation, so the momentum of the big body increases as in a random
walk.  After $N$ scatterings, the momentum of the big body grows to
$N^{1/2}mu$. The speed $v$ of the big body doubles when $N^{1/2}mu\sim
Mv$, i.e., after $N\sim (Mv/mu)^2$ scatterings.  Thus
\begin{eqnarray}
{1\over v}{dv\over dt}\Big{\vert}_{\rm df\ heat}
&\sim&
n_suR^2 \Big({mu\over Mv} \Big)^2 
\label{eq:heat}
\\
&=&
-{mu^2\over Mv^2} \cdot
{1\over v}{dv\over dt}\Big{\vert}_{\rm df\ cool} 
\ .
\end{eqnarray}
Hence $dv/dt\vert_{\rm df\ cool}+ dv/dt\vert_{\rm df\ heat}=0$ when
$Mv^2=mu^2$. As expected 
from thermodynamics,
dynamical friction drives the big body to equipartition with the small
bodies.

In most of our applications, big bodies have higher energies than do small
ones so dynamical friction cooling dominates dynamical friction
heating.

\subsubsection{$du/dt$ due to Dynamical Friction Cooling and Heating}
\label{subsubsec:velubigdfu}

A small body is heated because it gains more energy
in head-on collisions than it loses in tail-on ones.
It gains energy in head-on collisions
at the rate $du^2/dt\vert_{\rm head-on}\sim n_b R^2(u+v)[(u+2v)^2-u^2]$,
where $n_b$ is the number density of big bodies. It loses
energy in tail-on collisions at the rate
$du^2/dt\vert_{\rm tail-on}\sim n_bR^2(u-v)[(u-2v)^2-u^2]$.
Adding the two rates, we get
\begin{equation}
{1\over u}{du\over dt}\Big{\vert}_{\rm df \ heat}
\sim
 n_b {v^2\over u} R^2  \ 
\label{eq:heatu}
\end{equation}
to lowest order in $v/u$.

A small body is cooled by a sea of big bodies that
are nearly stationary ($v\sim 0$) because the small body loses energy
in each collision owing to the recoil of the big body. 
Because the big body recoils by $\Delta v\sim um/M$, its energy gain is
$\sim M(\Delta v)^2\sim (um)^2/M$, which is equal to the energy
lost by the small body. 
Thus $mdu^2/dt\vert_{\rm df\ cool} \sim -n_buR^2(um)^2/M$, or
\begin{equation}
{1\over u}{du\over dt}\Big{\vert}_{\rm df \ cool}
\sim
-n_b  uR^2{m\over M}  \ .
\label{eq:coolu}
\end{equation}

Because dynamical friction conserves energy, we could also have used
detailed balance,
\begin{equation} 
mn_s{du^2\over dt}\Big{\vert}_{\rm df \ heat\ or\ cool}= 
-Mn_b{dv^2\over dt}\Big{\vert}_{\rm df \ cool\ or\ heat} \ ,
\end{equation}
to derive Equations \ref{eq:heatu} and 
\ref{eq:coolu}.  
However, this approach is less physically enlightening.

\subsubsection{Viscous Stirring}
\label{subsubsec:velubigvs}

Random kinetic energy is not conserved in a circumsolar disk.  Viscous stirring tends to increase the random energies
of all bodies. 
This is similar to the way a viscous fluid undergoing shear converts free energy
associated with the shear into thermal energy through viscosity.
  In a circumstellar disk, the shear,$a|d\Omega/da|=3\Omega/2$, is Keplerian.
However, for protoplanets and planetesimals the fluid analogy is
imprecise because the collision rates are much smaller than the rate
of shear.

We consider first how a sea of big bodies moving on circular orbits
viscously stirs a small body as it travels around the Sun.  Relative
to a circular orbit, the small body's azimuthal and radial speed are
of order $u$.  An elastic collision rotates the relative velocity
vector while preserving its length.\footnote{Here we neglect the
recoil of the big body, which is accounted for by the dynamical friction
heating of the big body and cooling of the small one.}  That such
rotation tends to increase the random energy of the small body can be
understood as follows (\citen{S72}):  A
body's speed relative to a cospatial particle moving on a circular
orbit is twice as large at quadrature as it is at either periapse or
apoapse. Collisions that occur when the small body is at quadrature,
and rotate its orbit so 
that
immediately after the collision
 the small body is
at periapse or apoapse,
double the
epicyclic amplitude. Those collisions that occur when the small body is at
periapse or apoapse and rotate its orbit so that the small body is at
quadrature halve the epicyclic amplitude. Thus the former increase the
random kinetic energy by a factor of four and the latter reduce it by
the same factor. 
Safronov (1972) assumed an equal number of collisions in each direction, which
leads
 to an average increase in random kinetic energy per collision by
the factor $(4+1/4)/2=17/8$. 
His assumption, while quantitatively imprecise, captures the essence of viscous
stirring, and implies
that
the heating rate from viscous stirring is similar to the collision
rate:
\begin{equation}
{1\over u}{du\over dt}\Big\vert_{\rm vs}\sim
n_buR^2 \ .
\label{eq:viscousu}
\end{equation}

Similarly, a big body is viscously stirred by a sea of small bodies in
the time it takes for collisions to deflect the big body's epicyclic velocity by an
order-unity angle away from that body's unperturbed motion. 
 Because this time
is the same as that required for $v$ to double by dynamical friction
heating, the viscous stirring rate for the big body is the same as its
dynamical friction heating rate 

\begin{equation}
{1\over v}{dv\over dt}\Big\vert_{\rm vs} \sim
{1\over v}{dv\over dt}\Big\vert_{\rm df\ heat}
\sim n_suR^2\Big({mu\over Mv}\Big)^2 \ ,
\label{eq:viscousv}
\end{equation}
where the last equality follows from Equation \ref{eq:heat}.

\subsubsection{Collected Results for Elastic Collisions with $u>\vesc$}
\label{subsubsec:collecting}

The big bodies' velocity dispersion, $v$, is affected by the small bodies
through dynamical friction cooling
(Equation  \ref{eq:simplecool}) and
heating (Equation \ref{eq:heat}), and viscous stirring (Equation \ref{eq:viscousv}):
\begin{eqnarray}
{1\over v}{dv\over dt}\Big\vert_{\rm df\ cool}
&\sim&-\Omega {\sigma\over\rho R} \label{eq:vel1} \\
{1\over v}{dv\over dt}\Big\vert_{\rm df\ heat}\sim {1\over v}{dv\over dt}\Big\vert_{\rm vs}
&\sim&\Omega {\sigma\over\rho R}{mu^2\over Mv^2}\label{eq:vel2} \ ,
\end{eqnarray}
where we have replaced $n_s$ with $\sigma\Omega/(mu)$.  Cooling
acts in the time that it takes the big body to collide with a mass $M$ of small
bodies. The time needed by heating and viscous stirring to act is longer by
$Mv^2/mu^2$. 

The small bodies' velocity dispersion, $u$, is significantly affected
by the big bodies only through viscous stirring
(Equation \ref{eq:viscousu}), which acts in the time that it takes a
small body to collide with a big body:
\begin{equation}
{1\over u}{du\over dt}\Big\vert_{\rm vs}
\sim \Omega{\Sigma\over\rho R} \ ,
\label{eq:vel3}
\end{equation}
where we have replaced $n_b$ with $\Sigma\Omega/(Mu)$.  Although the number density of big bodies in the midplane is
$\Sigma\Omega/(Mv)$, the scale height of the small bodies is
larger than that of the big bodies by $u/v$.  Thus the effective number
density of big bodies, when interactions with small bodies are
involved, is smaller than their midplane number density by $v/u$.
Heating and cooling 
of $u$
by dynamical friction
(Equations \ref{eq:heatu} and \ref{eq:coolu}) are less important than
viscous stirring by the factors $v^2/u^2<1$ and $m/M<1$, respectively,
so it is safe to neglect them.

\subsection{Velocity Evolution when $\vesc>u>v_H$}
\label{subsec:subescape}

For $u<\vesc$, gravitational focusing is important. The principal
exchange of momentum between big and small bodies is by collisionless
gravitational deflections. A small body that passes within a distance
$\sim GM/u^2\sim R(\vesc/u)^2$ of the big body is deflected by an
angle of order unity. Therefore the effective cross section for momentum 
exchange is $R^2(\vesc/u)^4$, not $R^2$. Appropriate rates are
obtained by multiplying Equations 24--26 by
$(\vesc/u)^4$:

\begin{equation}
{1\over v}{dv\over dt}\Big\vert_{\rm df\ cool} \sim -\Omega
{\sigma\over\rho R}\left({\vesc\over u}\right)^4\,\, ,
\label{eq:chandra}
\end{equation}
\begin{equation}
{1\over v}{dv\over dt}\Big\vert_{\rm df\ heat}\sim {1\over v}{dv\over
dt}\Big\vert_{\rm vs} \sim \Omega {\sigma\over\rho R}{m\over
M}\left(v_{\rm esc}^2\over u\, v\right)^2 \,\, ,
\label{eq:vheat}
\end{equation}
\begin{equation}
{1\over u}{du\over dt}\Big\vert_{\rm vs}\sim \Omega
{\Sigma\over\rho R}\left({\vesc\over u}\right)^4 \,\, 
\label{eq:uvs}
\end{equation}
(see Appendix D for a more complete derivation of Equation 27). As noted in Section \ref{subsubsec:collecting}, we can neglect dynamical friction
by big bodies on small bodies. 

\subsection{Velocity Evolution when $u<v_H$}
\label{subsec:hillvelocity}

Small bodies enter the Hill sphere of a given big body at the rate
given by Equation \ref{eq:hillentryrate} and exit with random
velocities $\sim v_H$. A big body is cooled by dynamical friction when it
strongly scatters a mass $M$ of small bodies (see Section 4.1.1):

\begin{equation}
{1\over v}{dv\over dt}\Big{\vert}_{\rm df\ cool}
\sim 
-{\sigma\over m}\Omega R_H^2 {m\over M}
\sim
-\Omega {\sigma\over\rho R}
{1\over\alpha^2} \ .
\label{eq:vcoolultvh}
\end{equation}

A big body is heated by viscous stirring when small bodies deflect its
random velocity by an angle of order unity.  Provided
$mv_H<Mv$, 
this occurs after $N\sim (Mv/mv_H)^2$ small bodies enter the big body's Hill sphere (see Section
\ref{subsubsec:velubigdfv}). Therefore
\begin{equation}
{1\over v}{dv\over dt}\Big\vert_{\rm vs} \sim {\sigma\over
m}\Omega R_H^2\Big({mv_H\over Mv}\Big)^2 \sim
\Omega{\sigma\over\rho R}{1\over \alpha^2}{mv_H^2\over
Mv^2} \ .
\label{eq:vvsultvh}
\end{equation}

Out of the total momentum transfered in a strong scattering of a small
body by a big one, $\sim mv_H$ contributes to viscous stirring of the
big body and $mu$ to its dynamical friction heating. Thus the
dynamical friction heating rate is smaller than the viscous stirring
rate by the factor $(u/v_H)^2$ and can be neglected. This result is
peculiar to the case $u<v_H$. For $u>v_H$, the rates of dynamical
friction heating and viscous stirring are equal, independent of whether
$u<\vesc$ or $u>\vesc$.

Viscous stirring of small bodies for $u<v_H$ involves subtleties that have gone
unrecognized until recently (Rafikov 2003e).  For all other processes that we consider, it suffices to leave
the definition of $u$ imprecise, but for viscous stirring with $u<v_H$ it does not.
Consider the case that the velocities of all small bodies are much less than $v_H$.
Then after only a few small bodies have entered the Hill spheres of big bodies---boosting
those small bodies' energies to $mv_H^2$---there will be a substantial increase in $\overline{u^2}$. But
the random speed of a typical small body will not have changed. 
Hence one must specify whether it is 
the rms speed 
$(\overline{u^2})^{1/2}$ or the speed of a typical 
body that is of interest.
 Most applications
require the typical speed of a small body, not the rms speed.\footnote{
More precisely, 
the typical speed regulates the rates of accretion and dynamical friction, whereas
the rms speed enters in the time-averaged rate of viscous diffusion.}

We use $u$ to denote the typical speed.  The rate at which $u$ increases
is then equal to the rate at which the speed of a typical small body is doubled.
If $u\ll v_H$, a small body that enters a big body's Hill sphere has its random energy more
than doubled.  Thus more distant interactions
 suffice to double $u$.
Because encounters with a big body at the
impact parameter,
\begin{equation}
b_{\rm scat}\sim \left(v_H\over u\right)^{1/2}R_H\,\, ,
\label{eq:bstir}
\end{equation}
rotate a random velocity vector of magnitude $u$ by a large angle, the viscous stirring rate is
\begin{equation}
{1\over u}{du\over dt}\Big\vert_{\rm vs}\sim{\Sigma\over
M}\Omega b_{\rm scat}^2 \sim\Omega{\Sigma\over\rho R}{1\over
\alpha^2} {v_H\over u} \, \, .
\label{eq:vssubhill}
\end{equation}

We can estimate the rate at which $\overline{u^2}$ increases by  taking the product of
two terms:
({\it a}) the rate
at which small bodies enter the big bodies' Hill spheres and ({\it b}) the
energy boost  that a small body receives upon exiting
a big body's Hill sphere ($mv_H^2$).
After dividing by $\overline{u^2}$ to obtain the logarithmic rate, this product yields

\begin{equation}
{1\over {\overline {u^2}}}{d{\overline {u^2}}\over dt}\Big\vert_{\rm
vs}\sim {\Sigma\over M}\Omega R^2_H{v_H^2\over {\overline
{u^2}}}\sim \Omega{\Sigma\over \rho R}{1\over
\alpha^2}{v_H^2\over {\overline {u^2}}} \,\, .
\label{eq:vslitsmall}
\end{equation}
This is how viscous stirring is calculated in the literature, where 
$\overline{u^2}\sim u^2$ is also assumed
(e.g., Goldreich, Lithwick \& Sari 2002; Ida 1990; 
Weidenschilling et al. 1997; Wetherill \& Stewart 1993).
But this is incorrect because $\overline{u^2}>u^2$.

Next we deduce the distribution function for the velocities of the
small bodies. A small body approaches a big one with an impact parameter
$b>R_H$ at a relative speed $\sim \Omega b$. Because the duration
of the big body's peak gravitational acceleration, $\sim GM/b^2$, is
$\sim \Omega^{-1}$ (independent of $b$), the small body receives
a velocity kick $\Delta u\sim (R_H/b)^2v_H$. Because encounters
with the impact parameter $b$ occur at a rate proportional to $b^2$,
velocity kicks of order $\Delta u$ occur at a frequency
proportional to $(\Delta u)^{-1}$.
The distribution function of random velocities mirrors that of the
velocity kicks. It consists of a peak at $\sim u$, which contains approximately
half the bodies, and a power law tail of slope minus two at higher 
velocities.
Only a small fraction ($\sim u/v_H$) of the bodies have velocities 
of order $v_H$, above which the power law is truncated.
The most probable and median velocities are both $\sim u$, but the 
major contribution to the rms velocity comes from
near the upper cutoff, $v_H$. The rms velocity is related to $u$ by
\begin{equation}
\left(\overline {u^2}\right)^{1/2}\sim \left(u v_H\right)^{1/2}\,\, ,
\label{eq:msvel}
\end{equation}
which implies that the rates of
viscous stirring for $\overline {u^2}$ and $u$ are comparable.

\subsection{Inclinations and Eccentricities Can Differ}
\label{subsec:inclinations}

Thus far, our derivations have included the implicit assumption that
inclinations and eccentricities have comparable magnitudes. Here we
relax this assumption and compare the rates of growth of inclination
and eccentricity during viscous stirring. Our concern is with typical,
as opposed to rms, values of inclination and eccentricity in the sense
described in Section \ref{subsec:hillvelocity}. Working with the vertical and
total components of the random velocity, $u_\perp$ and $u$, we assume $u_\perp<u$ and
compare the excitation rates of each. We then determine whether the growth of $u_\perp$ is able to keep pace with
that of $u$.

\subsubsection{$u<v_H$} 

As shown in Section \ref{subsec:hillvelocity}, viscous stirring of $u$ is
most effectively accomplished by scatterings at impact parameter
$b_{\rm scat}$ (Equation \ref{eq:bstir}). On average, each scattering approximately doubles $u$. Because vertical oscillations of
the small body carry it a distance of order $u_\perp/\Omega$ away from
the orbit plane of the large body, these scatterings produce vertical
impulses of magnitude $\Delta u_\perp$, where
\begin{equation}
{\Delta u_\perp\over u_\perp}\sim {u\over \Omega b_{\rm scat}} \sim
\left(u\over v_H\right)^{3/2}< 1 \,\, .
\end{equation}

Doubling of $u_\perp$ is accomplished most rapidly by closer
encounters with $b\sim R_H$. Because the scattering rate varies as
$b^2$, the growth rate of $u_\perp$ is slower than that of $u$
according to
\begin{equation}
{1\over u_\perp}{du_\perp\over dt}\sim \left(u\over v_H\right){1\over u}{du\over
dt}\,\, .
\end{equation}

\subsubsection{$u>v_H$}

Strong scatterings for $u>v_H$ occur at the impact parameter
\begin{equation}
b_{\rm scat}\sim \left(v_H\over u\right)^2 R_H\,\, .
\end{equation}
On average, each scattering approximately doubles $u$. Their effect on $u_\perp$
differs according to whether the vertical excursion $u_\perp/\Omega$ is
less than or greater than $b_{\rm scat}$.

In the former case, the scattering takes place when the vertical
displacement of the small body is of order $u_\perp/\Omega$: 
\begin{equation}
{\Delta u_\perp\over u_\perp}\sim {u\over \Omega b_{\rm scat}}\sim
\left(u\over v_H\right)^3>1\,\, .
\end{equation}
In the latter case, the small body's vertical displacement 
when it scatters
is of
the same order as $b_{\rm scat}$ so 
\begin{equation}
{\Delta u_\perp\over u_\perp}\sim {u\over u_\perp}>1\,\, .
\end{equation}
In both cases, strong scatterings result in $\Delta
u_\perp/u_\perp>1$. This implies that the growth of $u_\perp$ is
dominated by more frequent but weaker scatterings.   Thus for $u>v_H$, the growth of
$u_\perp$ is faster than that of $u$, provided $u_\perp<u$.

\section{Mass and Velocity Evolution: Summary and Extensions}
\label{sec:massgrowthrate}

\subsection{Collected Equations for $u<v_{\rm esc}$}
\label{sec:collectedequations}

A big body accretes small ones that suffer inelastic collisions with
it at $u<v_{\rm esc}$. The big body's mass growth rate is $m$ times the
collision rate.  Collecting Equations  \ref{eq:cr2}, \ref{eq:cr3},
and \ref{eq:cr4}, we arrive at
\begin{equation}
{1\over M}{dM\over dt} \sim {1\over R}{dR\over dt} \sim \Omega
{\sigma\over \rho R} F^{col} \left( u\over v_{esc}\right),
\label{eq:masslast}
\end{equation}
where
\begin{equation}
F^{col}(x)=
\cases{ 
        x^{-2}       \ ,& {$1>x>\alpha^{1/2}$} \cr
        \alpha^{-1/2}x^{-1} \ , & {$\alpha^{1/2}>x>\alpha$} \cr
        \alpha^{-3/2}    \ , & {$\alpha>x$} .
      }
\label{eq:Fcol}
\end{equation}

Small bodies are viscously stirred by big ones at the rate 
(Equations \ref{eq:uvs} and \ref{eq:vssubhill})
\begin{equation}
{1\over u}{du\over dt}\Big\vert_{\rm vs}
\sim \Omega{\Sigma\over \rho R} 
F^{vs}
\left( u\over v_{esc}\right),
\label{eq:vsfinala}
\end{equation}
where
\begin{equation}
\label{eq:Fvs}
F^{vs}(x)=
\cases{ 
        x^{-4} \ ,& {$1>x>\alpha^{1/2}$} \cr
        \alpha^{-3/2} x^{-1} \ , & {$\alpha^{1/2}>x$} .
      }
\label{eq:vslast}
\end{equation}
Dynamical friction heating and cooling of $u$ by big bodies is always
negligible relative to viscous stirring.

Big bodies are cooled by small ones at the rate
(Equations
\ref{eq:chandra} and \ref{eq:vcoolultvh})
\begin{equation}
{1\over v}{dv\over dt}\Big\vert_{\rm df\ cool}
\sim -\Omega {\sigma\over\rho R} F^{df}
\left( 
{u\over v_{\rm esc}}
\right),
\label{eq:coollast}
\end{equation}
where
\begin{equation}
F^{df}(x)=
\cases{ 
        x^{-4} \ ,& {$1>x>\alpha^{1/2}$} \cr
        \alpha^{-2} \ , & {$\alpha^{1/2}>x$} .
      }
\label{eq:Fdf}
\end{equation}
They are viscously stirred by small ones at the rate (Equations \ref{eq:vheat}
and \ref{eq:vvsultvh})
\begin{equation}
{1\over v}{dv\over dt}\Big\vert_{\rm vs}
\sim \Omega {\sigma\over\rho R} 
{mu^2\over Mv^2}
\cases{
        (u/v_{\rm esc})^{-4} \ ,& {$1>u/v_{\rm esc}>\alpha^{1/2}$} \cr
        \alpha^{-1}(u/v_{\rm esc})^{-2} \ , & {$\alpha^{1/2}>u/v_{\rm esc}$} .
      }
\label{eq:vsfinal}
\end{equation}
For applications in this review, viscous stirring by big bodies 
dominates that by small bodies.  In general, it is always safe to neglect
dynamical friction heating of big bodies by small ones. When $u>v_H$,
dynamical friction heating is comparable to viscous stirring; when $u<v_H$, it is less than
viscous stirring. 

In deriving these equations, we made the following main assumptions: $v<u$, $mu<Mv$, $u<a\Omega$, $u<v_{\rm esc}$, 
and inclinations are comparable to eccentricities.
We also only considered how big bodies affect small ones, and vice versa, but not how
either group affects itself.
In the following subsections, we extend these results to cover more general 
cases.

\subsection{How Bodies Affect Themselves}

\subsubsection{Interactions Among Big Bodies}
\label{sec:selfvs}

Expressions for interactions among big bodies are
obtained by replacing $u$ by $v$ and $\sigma$ by $\Sigma$.  Thus big
bodies coalesce at the rate (Equation \ref{eq:masslast})
\begin{equation}
{1\over R}{dR\over dt}\sim 
\Omega{\Sigma\over\rho R}
F^{col}\left({v\over v_{\rm esc}}\right) \ \ , \ \ \ \ \ v<v_{\rm esc}
\label{eq:selfgrowth}
\end{equation}
and stir themselves at the rate (Equation \ref{eq:vsfinala})\footnote{
The apparent discrepancy  between Equations
\ref{eq:vsfinal} and \ref{eq:selfstir} in the sub-Hill regime is a result of the assumption $mv_H<Mv$
that was used in deriving Equation \ref{eq:vsfinal}.}

\begin{equation}
{1\over v}{dv\over dt}\Big\vert_{\rm vs}\sim \Omega
{\Sigma\over\rho R} F^{vs}\left( {v\over v_{\rm esc}}\right)
\ \ , \ \ \ \ \ v<v_{\rm esc} \ . \label{eq:selfstir}
\end{equation}
Dynamical friction, in the collective
manner we use it, drives two groups to equipartition with each other; it
does not act within a single group of homogeneous bodies.

\subsubsection{Inelastic Collisions Among Small Bodies}

It
suffices to consider the case $u>u_{\rm esc}$ for which
gravitational focusing is unimportant.  
In Section \ref{subsec:velubig}, we
assume for pedagogical reasons that physical collisions are perfectly
elastic, in which case small bodies viscously stir themselves when
they collide. However, real collisions are inelastic.
They damp $u$ in a collision time:
\begin{equation}
{1\over u}{du\over dt}\sim -\Omega
{\sigma\over\rho s} \ \ , \ \ \ \ \ u>u_{\rm esc} \  , \label{eq:ucolfirst}
\end{equation}
where $s$ is the radius of a small body.
Inelastic collisions damp $u$ provided the coefficient of restitution 
falls below 
 a critical value (0.63 in the idealized model of Goldreich
\& Tremaine 1978).  Above the critical value, the net effect of collisions is to increase $u$.  When $u<u_{\rm esc}$, strong gravitational interactions
are more frequent than collisions, and they cause  $u$ to increase.
When $u\gg u_{\rm esc}$, collisions fragment small bodies that 
have little cohesive strength. We explore
the implications of fragmentation in subsequent sections. 

\subsection{Effect of Anisotropic Velocities on the Accretion 
Rate}
\label{subsec:anisotropy}

For 
$v<v_H$, 
the big bodies'
inclination growth 
by viscous stirring is
slower than 
their
eccentricity growth (see Section 4.4).  As 
discussed below, the velocity dispersion of large bodies is set by 
the balance between mutual eccentricity excitation and dynamical friction from small bodies.
Because dynamical friction acts on the horizontal 
and vertical
velocity with comparable timescales, inclinations are damped faster than they are
excited.
Large bodies with $v<v_H$ lie in a flat disk.
Thus the vertical thickness of the
disk of large bodies varies discontinuously across $v\sim v_H$. This
has one profound consequence: 
A flat disk enhances the focusing factor for
mutual collisions of large bodies across $v\sim v_H$.  So big bodies grow by accreting each other at a rate given
by inserting
\begin{equation}
F^{col}(x)=
\cases{
        x^{-2}       \ ,& {$1>x>\alpha^{1/2}$} \cr
        \alpha^{-3/2} \ , & {$x<\alpha^{1/2}$} \cr
      }
\label{eq:Fcolflat}
\end{equation}
 into Equation \ref{eq:selfgrowth}
(instead of inserting Equation \ref{eq:Fcol}).
  Similarly, collisionless 
small bodies with $u<v_H$ lie in a flat disk.  So big bodies grow by 
accreting them at a rate  given by inserting 
Equation \ref{eq:Fcolflat} for $F^{col}$
into Equation \ref{eq:masslast}.  But if small bodies collide frequently, they isotropize their velocity dispersion, in which case Equation \ref{eq:Fcol} for $F^{col}$ is applicable.

The flattening of a shear-dominated disk
 has 
 largely been
   neglected in the literature.
Wetherill \& Stewart (1993) and Weidenschilling et al. (1997)
discovered it in numerical simulations, and Rafikov (2003e) provided a simple explanation similar to ours.

\subsection{Mass Evolution for $u>v_{\rm esc}$}

Gravitational focusing does not operate when $u>v_{\rm esc}$, so the
collision cross section is $R^2$. An upper limit to the growth rate
of big bodies is obtained by adding the mass of the impacting small
bodies to that of the big body: Equation \ref{eq:cr1}
yields
\begin{equation}
{1\over R}{dR\over dt}\sim \Omega {\sigma\over\rho R} \ \ , \ \ \ \ \ u>v_{\rm esc} \ .
\label{eq:rdotbigu}
\end{equation}
However, fragmentation of the material composing the bodies may
result in net mass loss from the big body, especially if 
$u\gg v_{\rm esc}$.

\subsection{Physical Processes when $v>u$ \label{s:v>u}}

We consider first the case that $v_{\rm esc}>v>v_H$.  Because the
relative speed between big and small bodies is $v$, it is $v$ that
enters into the gravitational focusing factor for accretion. Thus
the accretion rate is given by Equation \ref{eq:masslast}
with $u$ replaced by $v$, i.e., $(1/M)dM/dt\sim (\sigma\Omega/\rho
R)(v_{\rm esc}/v)^2$.  For viscous stirring of small bodies by big
ones, a small body is deflected by $u$ when it passes
within a distance $R(v_{\rm esc}^2/uv)$ of a big one.\footnote{This
only holds if $u>v_H^3/v^2$.  Otherwise, at a distance $R(v_{\rm
esc}^2/uv)$ from the big body, the Keplerian shear velocity would
exceed $v$, in which case viscous stirring of $u$ would be given by
the formula appropriate for $v<v_H$.}  So the
cross section for viscous stirring differs from that used in Equation
\ref{eq:vsfinala} by the multiplicative factor $u^2/v^2$, and
$(1/u)du/dt\vert_{\rm vs}\sim (\Sigma\Omega/\rho R)(v_{\rm
esc}^4/u^2v^2)$.  This is comparable to the rate at which $u$ is
heated by big bodies via dynamical friction.  Cooling of $u$ by
dynamical friction with big bodies is negligible, because a big body has
more random energy than a small one.  Big bodies are cooled by
dynamical friction with the small bodies at a rate 
obtained by replacing $u$ with $v$
in Equation \ref{eq:coollast}:
$(1/v)dv/dt\vert_{\rm df\ cool} \sim -(\sigma\Omega/\rho
R)(v_{\rm esc}/v)^4$.

When $v<v_H$, the big bodies lie in a flat disk (see Section
\ref{subsec:anisotropy}).  If the small bodies were collisionless,
their disk would also be flat.  We consider instead the more practical
situation in which collisions roughly isotropize their velocity
distribution. Then there are no differences in the important rates
of accretion, viscous stirring, and dynamical friction between the
case $u<v<v_H$ and the one considered previously, i.e., $v<u<v_H$.  In
both cases, the rate at which big and small bodies approach each
other is independent of $u$ or $v$ and set by the Keplerian shear.
The mass accretion rate is given by Equation \ref{eq:masslast},
viscous stirring of small bodies is given by Equation
\ref{eq:vsfinala}, and dynamical friction of big bodies is
given by Equation \ref{eq:coollast}.  The accretion
rate is discontinuous across $v=v_H$ for $u<v_H$.

\subsection{Derivations of Mass and Velocity Evolution: Some Highlights}

The book by Safronov (1972)
 contains the first systematic
treatment of mass and velocity evolution in protoplanetary disks.  It
provides approximate expressions for mass evolution, viscous
stirring, and dynamical friction drag, but primarily for $\vesc>u>v_H$. Greenberg et al. (1991) derived analytic formulae for mass accretion
for $u<v_H$.
Dones \& Tremaine (1993) found order-unity coefficients when $u=0$ by performing
numerical integrations.

Numerous authors have employed increasingly sophisticated techniques in
more precise derivations of the velocity evolution rates for $\vesc>u>v_H$.
There are two types of derivations: one based on a Fokker-Planck equation
(the local-velocity formalism), and the other based on averaging
over orbital elements (the Hill velocity formalism).  Stewart \& Ida (2000) 
showed that both formalisms lead to the same results. In the process, they
corrected various order-unity errors that had appeared in previous
derivations.

Ida (1990) considered velocity evolution when $u<v_H$. He used
scaling arguments, as well as numerical experiments, for the case $u=0$
to estimate order-unity coefficients. Some of his
``order-unity'' coefficients are quite large---nearly $10^2$.
Ohtsuki, Stewart \& Ida (2002)\nocite{Oht02} compared predictions
of analytically derived formulas for dynamical friction and viscous stirring
with numerical integrations.

\section{Growth of Planets: An Overview}
\label{sec:growthofplanets}
Before solving the mass and velocity evolution equations, we present
general arguments about how planets might have formed. These lead us
to deduce that Uranus and Neptune grew by accreting small bodies.

\subsection{Observational Constraints}
\label{sec:observational}

In contemplating the formation of Uranus and Neptune, we are guided by
a few observational constraints.  Foremost among them are the planets'
masses and radii and the age of the Solar System. Gravitational
oblatenesses provide more subtle clues. Interior models that fit these
data probably require that Uranus and Neptune each contains 
a few Earth masses of hydrogen
and helium \cite{Guillot99}. This would imply that protoplanets
sufficiently massive to accrete an envelope of gas must have formed
prior to the total
loss of the Sun's protoplanetary gas disk. Observations
of young, solar-type stars suggest that circumstellar disks dissipate
over a timescale of several million years 
({Brice{\~ n}o}, {Vivas} \& {Calvet} 2001; Haisch, Lada \& Lada 2001\nocite{HLL01};
\citen{H02}; Lagrange, Backman \& Artymowicz 2000\nocite{LBA00};
Strom, Edwards \& Skrutskie 1993\nocite{Strom93}).

We ignore other observational clues that deserve
further attention. 
Among these are 
Uranus's and Neptune's
residual heat
(Hubbard, Podolak \& Stevenson 1996\nocite{HPS96}; Podolak, Hubbard \& Stevenson 1991\nocite{PHS91}), as well as their substantial obliquities and nonzero orbital eccentricities
and inclinations.  
Giant impacts are obvious candidates for producing the obliquities, but
they are not required:
At least for Saturn, a good case can be made for spin-orbit resonances
(Hamilton \& Ward 2002).

\subsection{Minimum Mass Solar Nebula}
\label{subsec:MMSN}

We adhere to the prevailing view that planets are born in disks around
young stars. The minimum mass solar nebula is a useful reference
disk \cite{Edgeworth49, Kuiper56, Weiden77, Hayashi81}. Its surface density of condensates is
\begin{equation}
\sigma_{\rm MMSN}\sim \left({a\over 10{\rm\
AU}}\right)^{-3/2}\gm\cm^{-2} \ \ , \ \ a\gtrsim 2.7 {\rm \ AU} \ \ .
\label{eq:mmsn}
\end{equation}
Hayashi (1981) arrived at this expression by imagining that the
material in each planet, other than hydrogen and helium, was
distributed in an annulus around the planet, such that the annuli of
neighboring planets just touch. The minimum mass solar nebula (MMSN) assumes Jupiter, Saturn, Uranus, and Neptune each
contributes 15
Earth masses, a
value that is consistent with modern interior
models \cite{GuillotGladman2000}. 
In our numerical expressions, we use the radius
\begin{equation}
R_{\rm planet}\sim 25,000 \km 
\label{eq:rplanet}
\end{equation}
for Uranus and Neptune and for the cores of Saturn and Jupiter.

Throughout this review, our default option is to assume that the
surface density of solid bodies is given by the MMSN, although it is
conceivable that, initially, it was much greater and that the excess
was either ejected from the Solar System or accreted by the Sun. A
higher initial surface density might shorten the timescale for planet
formation (e.g., Thommes, Duncan \& Levison 2003\nocite{Thommes03}). 
A possible measure of the initial surface density is given by the Oort cloud because
comets that reside there are thought to have formed near the planets, before
the planets kicked them out to their present large distances (more than a few thousand astronomical units).
The current mass of the Oort cloud is thought to be between 1 and 
50 Earth masses \cite{W96}.
If one also accounts for the mass that is completely ejected from the Solar System, the MMSN
might underestimate the original surface density 
by an order  of magnitude (Dones et al. 2001).

\subsection{Need for Gravitational Focusing}

If most of the time that it takes a planet to grow is spent
 in the last doubling of its mass, then its formation time
can be estimated from Equation \ref{eq:masslast}, after setting $(1/M)dM/dt\sim 1/t_{\rm form}$: 
 \begin{eqnarray}
 t_{\rm form}&\sim& \Omega^{-1} {\rho R\over \sigma } \left(
 {u\over v_{\rm esc}}\right)^2 \ \ \  \ \ v_H<u<v_{\rm esc} \ ,
 \label{eq:tform}
 \end{eqnarray}
where $u$ is the velocity dispersion of the mass that is
accreted during the final stages of its growth.\footnote{
Without gravitational focusing ($u\sim v_{\rm esc}$), and with $\sigma$ set by the MMSN, the formation
time is simply $t_{\rm form}\sim \Omega^{-1}a^2/R_{\rm planet}^2$ (the orbital
time divided by the planet's optical depth).}
Taking $\sigma$ to be
that of the MMSN (Equation \ref{eq:mmsn}) and $R$ to be the outer
planets' radii (Equation \ref{eq:rplanet}) gives
 \begin{eqnarray}
t_{\rm form}\sim 
15
\left({a\over 10 \au}\right)^3
\left({ u\over v_{\rm esc}}\right)^2 \ {\rm Gyr}
\label{eq:tformnumerically}
 \  .
\end{eqnarray} 
Without gravitational focusing, 
i.e., with $u=v_{\rm esc}$,
it would have taken Uranus and Neptune
100 Gyr and 400 Gyr, respectively, to form.  For Neptune to have formed in the
age of the Solar System, it must have accreted material with $u \lesssim v_{\rm esc}/10$. These estimates are based on
the assumption that the outer planets formed from the MMSN at their current
distances from the Sun.  An alternative scenario is given by
Thommes, Duncan \& Levison (1999, 2002)
\nocite{TDL99},
who suggest that Uranus and Neptune formed between 
 Jupiter and Saturn 
before Jupiter and Saturn accreted gas.

\subsection{Cooling Is Necessary, so Accreted Bodies Were Small}
\label{sec:smallbodies}
A planetary embryo viscously stirs all planetesimals that it significantly deflects.  But, provided $u<v_{\rm esc}$, only a small fraction of deflected planetesimals are accreted.
This is simply because the cross section for scattering exceeds that for accretion, i.e.,  
$F^{col}(u/v_{esc})<F^{vs}(u/v_{esc})$ when  $u<v_{esc}$.  Thus an embryo heats its food faster than it can eat it.
If the small bodies are not cooled by inelastic collisions or gas drag,
only a small fraction of them are accreted while $u<v_{\rm esc}$; most
are accreted when $u\sim v_{\rm esc}$.\footnote{More precisely, we show in Equation
\ref{eq:u} that 
$\Sigma/\sigma\sim (u/v_{\rm esc})^2$,
 where
$\Sigma$ and $\sigma$ are the surface densities of embryos and
planetesimals. $\Sigma$ only grows to be comparable to $\sigma$ when
$u\sim v_{\rm esc}$. 
In this section, we assume 
for simplicity
that
$v_{\rm esc}<\Omega a$.
Otherwise,
bodies would be ejected
from the Solar System before their speed reached $v_{\rm esc}$.} 
Therefore without cooling,  most planetesimals are accreted when gravitational focusing
is inoperative, and Uranus and Neptune would have taken far too long to
form.  
These general conclusions are in accord with
results from N-body simulations by Levison \& Stewart (2001) that follow the
evolution of a system initialized with a few hundred equal mass bodies
whose total mass slightly exceeds that of Uranus and Neptune
combined. These simulations, which also include perturbations from
Jupiter and Saturn, find almost no growth in the region of Uranus and
Neptune because viscous stirring is faster than accretion.
Levison, Lissauer \& Duncan (1998)
and Levison \& Stewart (2001) showed that earlier work arriving at contradictory 
conclusions \cite{Ip89,FI96,BF99} is incorrect.

How small must the bodies accreted by Neptune have been for them to be cooled by inelastic collisions?  Balancing
viscous stirring by an embryo (Equation \ref{eq:vsfinala} with
$\Sigma\sim\sigma$) with inelastic collisions between small bodies of radius $s$
(Equation \ref{eq:ucolfirst}) gives $s/R\sim (u/v_{\rm
esc})^4$.  Because $u\lesssim v_{\rm esc}/10$ is required
for Neptune to have formed in the age of the Solar System, 
the accreted bodies must have been smaller
than a few kilometers.\footnote{Cooling by
gas drag could not relax the upper limit on $s$
unless a cosmic abundance of gas were present for  at least a few times $10^8$ years.}
Rafikov (2003e) and Thommes, Duncan \& Levison (2003) found 
protoplanets grow rapidly when they accrete small bodies;
Rafikov (2003e) considers the damping of $u$ by gas drag.

 Although this review focuses on the formation of Uranus and Neptune, we mention briefly
 that for Earth,  Equation \ref{eq:tform} gives $t_{\rm form}\sim 0.1$ Gyr without gravitational focusing.  Hence Earth need not have accreted small bodies.  Indeed, the Moon is thought to have formed when Earth accreted a body not much smaller than itself \cite{CA01}.

\subsection{Completion: When Small Bodies Have Been Consumed}

If the planets formed out of the MMSN, they must have eaten
all the condensates in the disk.
Uranus and Neptune ate these condensates
in the form of bodies smaller than 1 km (Section \ref{sec:smallbodies}).
We refer to the epoch when 
all small bodies have been consumed as completion.
Any formation scenario that produces numerous subplanet-size big bodies beyond Saturn's orbit
at completion is unacceptable.  Although subplanets might have formed quickly, further significant growth is impossible in the age of the Solar System, because subplanets viscously stir each other
faster than they coalesce (see Section \ref{sec:smallbodies}). How could Uranus and Neptune have consumed all small bodies that were between them?
In the following, we consider two possibilities.

\subsubsection{Completion With $u\sim\Omega a$}
\label{sec:thesimplest}

 The simplest possibility is that the
orbits of accreted bodies crossed those of the growing protoplanets as
a result of their epicyclic motions.  For the accreted bodies'
epicycles to be as large as $a$, they must have had
$u/\Omega\sim a$.  Of course, $u$ could not have exceeded
$a\Omega$, or the bodies would have escaped from the Solar System.\footnote{$a\Omega/v_{\rm
esc}\sim 0.3 (a/10\au)^{-1/2}$ for $R=25,000\km$.}   
With $u\sim\Omega a$,
Equation \ref{eq:tformnumerically} gives
\begin{equation}
\label{eq:tforma}
t_{\rm form}\sim 2 \left({a\over 10 {\rm\ AU}}\right)^2
\ {\rm Gyr} \ ,
\end{equation}
i.e., $t_{\rm form}\sim 20$ Gyr for Neptune.
 Thus the giant planets could barely have formed
in the age of the Solar System if $u\sim \Omega a$.
But there is a serious problem with
this scenario.  
Small bodies must experience frequent
collisions (see Section \ref{sec:smallbodies}), and their sizes must be $\sim1$ km to keep $u\sim \Omega a$.  
These collisions would fragment them because
 $\Omega a\sim $ few kilometers per second  
 greatly exceeds the escape speed from their surfaces,  $u_{\rm esc}\sim 1$ m/s.
Fragmentation implies smaller bodies, which implies more effective inelastic damping,
and hence smaller $u$.  This leads us to examine the case $u\ll \Omega a$.

\subsubsection{Completion with $u\ll a \Omega$}
It is also possible that $u\ll a\Omega$.  The accretion
time could then easily be less than 10 Myr.  However, the epicycles of
the accreted bodies would only allow them to accrete onto nearby
planetary embryos.  For example, if $u\ll v_H$, an embryo could only
accrete from within its Hill sphere, i.e., within an annulus of
half-width $2.5 R_H$ (Greenberg 1991).\footnote{We retain factors of
order unity in the present calculation.}  The embryo's maximum radius
$R$ is given by $2\pi \sigma a (5 R_H)=(4/ 3)\pi\rho R^3$, where
$R_H\cong 0.62R (a/R_\odot)$ and taking $\rho\cong 1\gm\cm^{-3}$ for
the embryo. Solving for $R$ and setting $\sigma=\sigma_{\rm MMSN}$
yields the Hill isolation radius:
\begin{equation}
R_{\rm iso}\cong 12,000 \left( {a \over 10 \au}\right)^{1/4} \ \km \ 
\label{eq:isolationradius}
\end{equation}
(e.g., Lissauer 1993).  At 20--30 AU, the final
masses of the isolated embryos would be smaller than Uranus and
Neptune (Equation \ref{eq:rplanet}) by a factor of  $\sim5$.  So if
accretion with $u\ll v_H$ proceeded to completion, one would
expect approximately 10 small planets beyond Saturn.

This conclusion could be avoided if 
a mechanism other than the small bodies' 
epicyclic motions enabled them to travel a distance $a$.
For example, small
bodies might diffuse a distance $a$ owing to collisions among
themselves, or they could drift as a result of gas drag
(\citen{Weiden77b}; Kary, Lissauer \& Greenzweig 1993\nocite{KLG93}).
 Alternatively, the planets themselves might migrate
(Bryden, Lin \& Ida 2000\nocite{BLI00}; \citen{FIP84}; \citen{HM99};
\citen{TI99}; \citen{Ward89}).

An alternative that we regard as most likely is that $\sigma$ was a few times larger than $\sigma_{\rm MMSN}$. 
Because $R_{\rm iso}\propto\sigma^{1/2}$, this  
would have produced
planets whose radii are equal to those of
Uranus and Neptune.  But it would have produced half a dozen of them, instead of two.  We discuss what happens to the extra planets in Section \ref{s:after}.

We have yet to address how $u\ll v_H$ might be maintained. Although
cooling by gas drag is a possibility, we favor inelastic
collisions because it appears that the final assembly of Neptune took
place after most of the gas had dissipated. As seen by equating
Equations \ref{eq:vsfinala} and \ref{eq:ucolfirst}, collisional
cooling requires $s\ll R\alpha^2$. For Neptune, this translates
into $s<30$ cm. Such small bodies might result from fragmentation. 

\section{Solving the Evolution Equations}
\label{sec:solving}
\subsection{N-Body and Particle-in-a-Box Simulations}

In principle, one could initiate a simulation with many small bodies
orbiting the Sun and integrate the equations of
motion.  However, aside from fundamental
uncertainties such as the initial size of the bodies and whether
colliding bodies stick or fragment, this approach is computationally
impossible.  To describe the formation of 25,000-km bodies out of,
say, 1-km ones, requires following an astronomical number, $25,000^3$, of particles.
Nonetheless, N-body simulations can be useful for following the evolution
of a few hundred
bodies
(\citen{Cham01}; \citen{IM92}; Kokubo \& Ida 1996, 2000; \citen{LS01}).

An alternative to the N-body approach is to group all bodies with
similar masses together and to treat each group as a single entity
that has three properties: the number of bodies it contains, the mass
per body, and the characteristic random velocity.\footnote{More
sophisticated simulations differentiate between eccentricity and
inclination.}  Each group interacts with every other group (and with
itself); groups change each other's properties at rates that are given
by the formulae in Section \ref{sec:massgrowthrate}.  This
particle-in-a-box approach was pioneered by Safronov (1972) and
extended by many others (Davis et al. 1985; Greenberg et al. 1978, 1984; Inaba et al. 2001; Kenyon 2002; Kenyon \& Bromley 2001; Kenyon \& Luu 1998, 1999;
Ohtsuki, Nakagawa \& Nakazawa 1988; Spaute et al. 1991; Stern \& Colwell 1997; 
Weidenschilling et al. 1997; 
Wetherill \& Stewart 1989, 1993; Wetherill 1990).

Although particle-in-a-box simulations represent a great
simplification relative to N-body simulations, they are still
computationally expensive (Kenyon \& Luu 1998, for example, ran their simulations on a
CRAY supercomputer).  Moreover, the results of the simulations are very
difficult to interpret.  Because each group interacts with every other
group, it is not clear which groups are important for viscously
stirring the others, which are important for dynamical friction, or
whether bodies grow primarily by merging with others of comparable
size or by accreting much smaller ones.  
Disentangling these effects is important
for understanding planet
formation and for assessing how various uncertainties---such as
fragmentation and the initial sizes of the bodies---affect the final
results.
 
\subsection{Two-Groups Approximation}
\label{sec:twogroups}
A further simplification is to consider the evolution of only two
groups of bodies: big and small ones (e.g., \citen{WS89}, \citen{IM93}).  In the remainder of the review, we concentrate on this
two-groups approximation.  We use the same notation used in Section
\ref{sec:massgrowthrate} to refer to the properties of these two
groups ($\sigma, u, s, u_{\rm esc}, \Sigma, v, R, v_{\rm esc}, v_H$).
Motivated by the results of particle-in-a-box simulations, we define
the small bodies as those that both contain most of the mass and
provide most of the dynamical friction, and we define the big ones as
those that dominate the viscous stirring, even though they may
contribute very little to the total mass.  The identification of these
two groups with the results of numerical simulations depends on the
mass and velocity spectra.

We collect the interaction rates summarized in Section \ref{sec:massgrowthrate}.
Big bodies grow at the rate (Equations 
41 and 48)
\begin{equation}
\label{Rdot}
{1 \over R}{dR \over dt}\sim
\Omega{\Sigma  \over \rho R} F^{col}
\left(
{v\over v_{\rm esc}}
\right)+
\Omega{\sigma  \over \rho R} F^{col}
\left(
{u\over v_{\rm esc}}
\right) \ , \label{eq:rdotreem}
\end{equation}
where $F^{col}$ is given by Equation \ref{eq:Fcolflat}; however, if
 small bodies experience
frequent collisions that isotropize their random velocities, the 
second term should  instead use Equation \ref{eq:Fcol} for $F^{col}$.  Random velocities of
small and large bodies evolve at the rates (Equations 43, 45, 49, 50)
\begin{eqnarray}
\label{udot}
{1 \over u}{du \over d t}&\sim&
\Omega{\Sigma  \over \rho R} F^{vs}
\left({u\over v_{\rm esc}}\right)-
\Omega{\sigma  \over \rho s} 
\label{eq:udotreem}
\\
\label{vdot}
{1 \over v}{dv \over d t}&\sim&
\Omega{\Sigma  \over \rho R} F^{vs}
\left({v\over v_{\rm esc}}\right)-
\Omega{\sigma  \over \rho R} F^{df}\left({u\over v_{\rm esc}}\right) \ .
\label{eq:vdotreem}
\end{eqnarray}                    
In addition to the assumptions 
listed at the end of Section \ref{sec:collectedequations},
we also assume that $u>u_{\rm esc}$ and that viscous stirring of big
bodies by small ones is negligible.  Equations
\ref{eq:rdotreem}--\ref{eq:vdotreem} are three equations for five
unknowns, $R$, $u$, $v$, $s$, and $\Sigma$.  The quantities
$\Omega$ and $\rho$ are known, and $v_{\rm esc}\sim R
(G\rho)^{1/2}\sim \alpha^{-3}R\Omega$. Because initially
most of the mass is in small bodies, $\sigma$ does not evolve until
$\Sigma$ grows to a comparable value; in numerical estimates, $\sigma$ is set by the minimum mass solar nebula (see Section \ref{subsec:MMSN}).

The left-hand side of Equation \ref{eq:rdotreem} represents the inverse
of the time at which big bodies---i.e., those bodies that dominate the
stirring---have reached radius $R$.  One might expect that, if big
bodies grow by accreting small ones, then their number should not
evolve, i.e., $\Sigma\propto R^3$. Similarly, if big bodies grow by
accreting each other, one might expect that $\Sigma=$ constant. In
general, neither expectation is met.  
Even if big bodies accrete only each other, or only very
small bodies, different subsets of big bodies
can dominate the stirring, and thus $\Sigma$, at various times.

How then does $\Sigma$ grow?  In Sections \ref{sec:superhilloligarchy} and
 \ref{sec:subhilloligarchy}, we provide appropriate expressions for
$\Sigma$ during oligarchy, an important stage of planet formation.
However, at earlier times, $\Sigma$ must be obtained by solving an
integro-differential equation (the coagulation equation), which is a
difficult task \cite{Lee00, MG01}.  Although irrelevant for the final stages of planet formation, the value of $\Sigma$ at early
times is important for systems that have not yet reached oligarchy,
such as the Kuiper Belt, and for determining when oligarchy
begins. Numerical simulations have produced
contradictory results 
(Davis, Farinella \& Weidenschilling 1999; Inaba et al. 2001; Kenyon \& Luu 1998; Lee 2000; Wetherill \& Stewart 1993).
The
observed spectrum of Kuiper Belt objects
(Bernstein et al. 2003; Luu \& Jewitt 1998; 
Luu, Jewitt \& Trujillo 2000\nocite{LJT00};
\citen{TB01} ; 
Trujillo, Jewitt \& Luu 2001\nocite{TJL01})
may provide a clue, if, as
Kenyon \& Luu (1999) have proposed, it is a frozen image of the formation
epoch.  In the following, we treat $\Sigma$ as a free function. Where
appropriate, we insert the oligarchic expressions for $\Sigma$.

\subsection{Solutions in the Two-Groups Approximation}

\subsubsection{Overview of Method of Solution}

With the aid of physical insight, Equations
\ref{eq:rdotreem}--\ref{eq:vdotreem} can be reduced to algebraic
relations that are trivial to solve.  
Nonetheless, the algebra is messy, particularly because of the
multitude of cases.
 We outline how to solve the equations before 
actually solving them.
We treat $\Sigma$ as an unknown
function and assume $s$ is fixed, either at its initial value or
at some as-yet unspecified value.  There remain three equations for
three unknowns: $v$, $u$, and $R$.

On big bodies, viscous stirring is balanced by dynamical friction,
so $v$ is in quasi-steady state and is determined by equating the right hand side of Equation
\ref{eq:vdotreem} to zero.
Two possibilities exist for $u$.  If small bodies have had time to
collide with each other, then $u$ is in quasi-steady state: Viscous stirring of small bodies by big ones occurs at the same rate
as damping by inelastic collisions, so $u$ is given by setting the
right-hand side of Equation \ref{eq:udotreem} to zero.  If small
bodies have not yet collided, the second term on the right-hand side
of Equation \ref{eq:udotreem} is negligible. 
Because the remaining term is proportional to $u$ raised to some power, $u$ increases on
timescale $t$ such that $(1/u)du/dt\sim 1/t$
once $u$ has grown to be much larger than its initial value.
Equation \ref{eq:rdotreem} for $R$ can also be solved by replacing
$(1/R)dR/dt$ with $1/t$. 
We give some explicit solutions below,
focusing on a few
 that have the
virtues of relevance and simplicity.

\subsubsection{Quasi-Steady State of $v$}
\label{subsubsec:vsolution}

The two terms on the right-hand side of Equation \ref{eq:vdotreem}
are always larger than the corresponding terms on the right-hand side
of Equation \ref{eq:rdotreem}.  Viscous stirring acts in the time
that a big body is significantly deflected.  Similarly, dynamical
friction acts in the time that $M/m$ small bodies are significantly
deflected.  Because only a subset of the deflected bodies are accreted,
$v$ achieves a quasi-steady state on a timescale shorter than that at
which $R$ grows. Thus
\begin{equation}
\Sigma F^{vs}\left({v\over v_{\rm esc}}\right)
\sim \sigma F^{df}
\left({u\over v_{\rm esc}}
\right) \ 
\end{equation}
gives one relation between $u$, $v$, and $R$.

The evolution equation for $u$ (Equation \ref{eq:udotreem}) does not
involve $v$. However, $v$ is present in the first term on the right-hand side of the evolution equation for $R$ (Equation \ref{eq:rdotreem})
which arises from the merging of big bodies. Nevertheless, the value of
$v$ does not appear in the solution of Equation \ref{eq:rdotreem}.

If
$v>v_H$, then
\begin{equation}
v\sim u
\left(
{\Sigma\over\sigma}
\right)^{1/4} \  ,\ \  \ \ \  {v>v_H} \ \ 
\label{eq:vitou}
\end{equation}
and the growth of $R$ due to the accretion of small bodies proceeds
faster, by the factor $(\sigma/\Sigma)^{1/2}$, than that due to
merging of big bodies. 
By contrast, Makino et al. (1998) concluded that big bodies grow primarily by merging
with other big bodies.  But their result was derived under the assumption that
the velocities of all bodies follow from energy equipartition, whereas
we find a different result (see Equation \ref{eq:vitou} and Figures \ref{fig:spec}
and \ref{fig:spec2}). 

If $v<v_H$, then the disk of big bodies is
flat and big bodies accrete each other at the $v$-independent rate
$\alpha^{-3/2}\Sigma\Omega/(\rho R)$. Whether embryo
growth is dominated by the accretion of big or small bodies can be
determined by comparing the two terms in Equation \ref{eq:rdotreem}
once a solution for $u$ is at hand.

\subsubsection{Collisionless Solution with $u>v_H$ and $v>v_H$}
\label{subsubsec:collisionlesssoln}
This case is a useful vehicle for comparing our simple analytic
solutions with detailed numerical simulations done by
Kenyon \& Luu (1998, 1999) and Kenyon (2002). It may describe the late stages of accretion in
the Kuiper belt.\footnote{However, the collisional limit discussed in
Section \ref{subsubsec:collisionsoln} is almost certainly more relevant to
the late stages of planet formation.} The collisionless limit applies for $t<t_{\rm collide}$, where
\begin{equation}
t_{\rm collide}\sim\Omega^{-1}{\rho s\over\sigma}\sim 0.6 \left(
{a\over 10{\rm\ AU}} \right)^3 \left({s\over {\rm km}}\right)\Myr \ .
\label{eq:tcollide}
\end{equation} 
is the collision time between
small bodies (Equation \ref{eq:udotreem}).
We use the MMSN value of $\sigma$ 
 (Equation \ref{eq:mmsn}) in numerical expressions.
 When $t<t_{\rm collide}$,
the second term
on the right-hand side of Equation \ref{eq:udotreem} is negligible.
Because big bodies grow by accreting small ones (Section \ref{subsubsec:vsolution}), we
 drop the first term on the right-hand side of Equation \ref{eq:rdotreem}.
Setting $(1/R)dR/dt\sim (1/u)du/dt\sim 1/t$ in Equations
\ref{eq:rdotreem} and \ref{eq:udotreem} yields
\begin{eqnarray}
\label{eq:u}
{u\over v_{\rm esc}}&\sim&
\left(
{\Sigma\over\sigma}
\right)^{1/2} \\
t&\sim&
\Omega^{-1}
{\rho R\over\sigma}{\Sigma\over\sigma}
\label{eq:tR} \ .
\end{eqnarray}
Given $\Sigma/\sigma$ at time $t$, the above formulae determine $R$ and $u$.

The assumption of super-Hill velocities ($v>v_H$) constrains
\begin{equation}
{\Sigma\over\sigma}>\alpha^{2/3}\, ,
\label{eq:Sigoversig}
\end{equation}
(see
Equations \ref{eq:vitou} and \ref{eq:u}),
and the restriction to $t<t_{\rm collide}$ constrains
\begin{equation}
s>{\Sigma\over\sigma}R\, 
\label{eq:slowerxxx}
\end{equation}
(see Equations \ref{eq:tcollide} and \ref{eq:tR}).

\begin{figure}
\centerline{\epsfxsize=5.5in\epsfbox{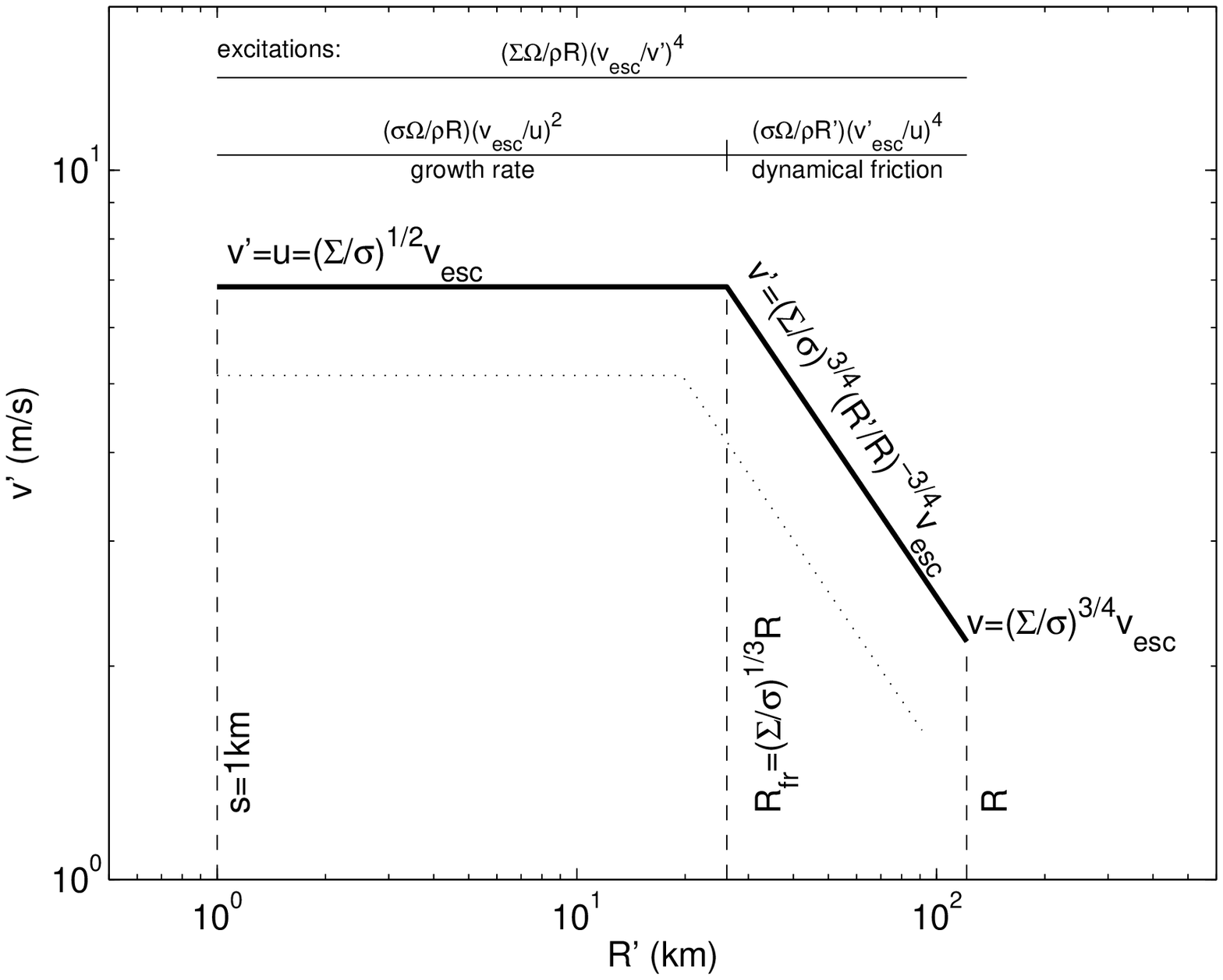}}
\caption{
Schematic plot of the random-velocity spectrum in the
collisionless case where all random velocities are in the dispersion-dominated regime.  The spectrum comprises two power-law segments
shown as bold solid lines. The rate at which viscous stirring by big
bodies excites random velocities is given above the upper horizontal
line. Rates that limit random velocities are written above the lower
horizontal line and identified immediately below it. Each power-law
segment of the spectrum is determined by equating the excitation rate
to the relevant limiting rate. Radii bounding each segment are written
along vertical dashed lines. Starting from the left, the first segment
describes viscous stirring limited by the growth time of the big
bodies, and the second viscous stirring is limited by dynamical
friction. Parameters chosen for this figure are $t\sim 20\Myr$,
$\sigma=0.2\gm\cm^{-2}$, $\Sigma=\sigma/100$, $a=35\au$,
$\rho=1\gm\cm^{-3}$, and $s=1\km$. For comparison, the dotted line
shows the spectrum at $t=15\Myr$ with all other parameters held fixed.
\label{fig:spec}}
\end{figure}

Next we determine the velocities of
intermediate-sized bodies (those with radii
$s<R^\prime<R$) (see Rafikov 2003d for a more detailed treatment). 
Bodies of intermediate size are passive; they
contribute negligibly to both viscous stirring and dynamical
friction.
The collisionless case is simplified by its mass-blind aspect: Up
to radius $R_{\rm fr}$, at which dynamical friction becomes
significant, the velocity dispersion is determined solely by viscous
stirring from big bodies. Thus the velocity
spectrum is flat between $s$ and $R_{\rm fr}$
(see Figure \ref{fig:spec}), where $R_{\rm fr}$
is the
radius at which dynamical friction damps the random velocity
$v^\prime$ on timescale $t$:
\begin{equation}
R_{\rm fr}\sim \left( \Sigma\over \sigma \right)^{1/3} R \, .
\label{eq:Rfr}
\end{equation} 
For consistency, $s$ must be smaller than $R_{\rm fr}$.

For $R_{\rm fr}<R^\prime<R$, viscous stirring by the big bodies and
dynamical friction from the small ones maintains the random velocity
$v^\prime$ in  quasi-equilibrium at 
\begin{equation}
v^\prime \sim v_{\rm esc} \left(R\over R^\prime\right)^{3/4}
\left(
{\Sigma\over\sigma}
\right)^{3/4}
\, .
\label{eq:vpvsRp}
\end{equation}
Although $v^\prime$ decreases with increasing $R^\prime$, the kinetic
energy increases proportional to 
$(R^\prime)^{3/2}$.

At this stage, we gather our analytic results and provide numerical
scalings that may be relevant to the formation of Kuiper belt
objects. Motivated by Kenyon's simulations, 
we choose the following
parameters: $\Sigma/\sigma=0.01$ and $\sigma=0.2\gm\cm^{-2}$ at
$a=35\au$. As demanded by Equation 67, $\Sigma/\sigma> \alpha^{2/3}$, so that $v>v_H$. From Equation
\ref{eq:tcollide}, we find that the collisionless approximation
applies up to $t_{\rm collide}\sim 20(s/\km)\Myr$.  Equation
\ref{eq:tR} determines the evolution of $R$:
\begin{equation}
R\sim {\sigma^2\Omega t\over\rho\Sigma}\sim 100\left(t\over
20\Myr\right)\km \, .
\label{eq:Revocl}
\end{equation}
From Equation \ref{eq:Rfr}, 
\begin{equation}
R_{\rm fr}\sim 0.2 R\sim 20\left(t\over 20\Myr\right)\km \, .
\end{equation}
The evolution equation for $u$ is obtained by combining Equations
\ref{eq:u} and \ref{eq:Revocl}:
\begin{equation}
u\sim \left(\Sigma\over\sigma\right)^{1/2}\vesc\sim 10\left(t\over
20\Myr\right)\m\s^{-1} \, .
\label{eq:uevocl}
\end{equation}
Taken together, Equations \ref{eq:vitou} and \ref{eq:uevocl} yield
the equation governing the evolution of $v$:
\begin{equation}
v\sim \left(\Sigma\over\sigma\right)^{3/4}\vesc\sim 3\left(t\over
20\Myr\right)\m\s^{-1} \, .
\label{eq:vevo}
\end{equation}
To facilitate comparisons between our analytic results and those obtained by
numerical simulations, we plot velocity dispersion as a function of
radius in Figure \ref{fig:spec}.  
We compare our Figure \ref{fig:spec} with  figure  10
from Kenyon \& Luu (1998), which yields input parameters for $\sigma$ and $\Sigma$
that our similar to ours at $t=20 \Myr$.
 Our simple approach reproduces the flat velocity
dispersion at $\cong 10\m\s^{-1}$ for bodies smaller than
$R^\prime\cong 20\km$. However, the velocity dispersion obtained by Kenyon \& Luu (1998)
declines more steeply 
above this transition than our analytic scaling
suggests it should.  We are unable to pinpoint the reason for this
discrepancy but note that it is ameliorated in later publications by the same group
(Kenyon \& Luu 1999, Kenyon 2002). 


Our approach captures some of the essential features
found by detailed simulations. 
In  Appendix B, we show how it can be extended to give  the velocity spectrum when
collisions are important.
However, our approach is incomplete because
we lack an evolution equation for $\Sigma$. This limitation is
removed during the oligarchic stage of planet growth during which
$\Sigma$ may be directly related to $R$
(Sections \ref{sec:superhilloligarchy}-\ref{sec:subhilloligarchy}).

\subsubsection{Collisional Solution}
\label{subsubsec:collisionsoln}

When collisions between small bodies are important, $u$ is damped by
collisions at the same rate as it is excited by viscous stirring,
i.e., Equation \ref{eq:udotreem} vanishes, so
\begin{eqnarray}
 {u\over v_H}&\sim& \alpha^{-2}{\Sigma\over\sigma}{s\over R}  \ \ \  \ \ , 
\ \ \ \  u<v_H
\\
{u\over v_H}&\sim&
\left(\alpha^{-2}
{\Sigma\over\sigma}{s\over R}
\right)^{1/4} 
\ \ \ , \  \ \ \ u>v_H 
 , \label{eq:ucolsecond}
\end{eqnarray}
after using $v_{\rm esc}\sim v_H\alpha^{-1/2}$.
For fixed $s$,  $u$ grows to super-Hill values at late times as long as $\Sigma$ increases
faster than $R$.

We select the case
 $v>v_H$ and, by implication, $u>v_H$, 
 in which to solve Equation \ref{eq:rdotreem} for $(1/R)dR/dt\sim 1/t$.
This case may be relevant at
the end of planet formation.

From Section \ref{subsubsec:vsolution}, big
bodies grow by accreting small ones, and Equation \ref{eq:rdotreem}
yields
\begin{equation}
t\sim
\Omega^{-1}
{\rho R\over\sigma}
\left(
{\Sigma\over \sigma}{s\over R}
\right)^{1/2}  \ .
\label{eq:tcoll}
\end{equation}
There are two restrictions on $R/s$ and $\Sigma/\sigma$:
\begin{eqnarray}
v>v_H &\Rightarrow& {R\over s}<\alpha^{-2}
\left({\Sigma\over\sigma}\right)^2 \\
t>t_{\rm collide} &\Rightarrow& {R\over s}>{\sigma\over \Sigma} \ .
\end{eqnarray} 
In this regime, it is always true that $u>u_{\rm esc}$ and $v<v_{\rm
esc}$.

\section{Orderly, Neutral, and Runaway Growth}
\label{sec:orderlyneutral}
The distribution function for the radii of large bodies can evolve in
three qualitatively different ways:  In orderly growth, the growth
rate $(1/R')dR'/dt$ 
at a fixed time
decreases with increasing $R'$.\footnote{Primes
attached to $R^\prime$, $\vesc^\prime$, and $v_H^\prime$ denote that
these symbols are being used for bodies with a range of radii. We
maintain our practice of reserving unprimed symbols for the big bodies
that dominate viscous stirring.}  Thus as big bodies grow, their radii
converge, and the radius distribution function
becomes progressively narrower.  In runaway growth, $(1/R')dR'/dt$ increases with
increasing $R'$.  Therefore the radius of the largest body increases
fastest and runs away from the radii of all smaller bodies. The
extreme high-$R'$ end of the distribution develops a tail to infinity
(Safronov 1972, Lee 2000, Malyshkin \& Goodman 2001).
 In neutral growth, $(1/R')dR'/dt$ is
independent of $R'$.  The radii of all bodies grow at the same rate,
so the distribution function of relative radii does not evolve.

From Equation \ref{eq:masslast} with $v_{\rm esc}'\propto R'$,
\begin{equation}
\label{eq:runaway}
{1 \over R'} {dR' \over dt} \propto \cases {R'^{-1}
\Rightarrow {\rm orderly}
 & $u>v_{\rm esc}'$
\cr 
 R'\Rightarrow {\rm runaway} & $v'_{\rm esc}>u>v'_H$
\cr
 1\Rightarrow {\rm neutral} &
$v_H'>u>\alpha^{1/2} v_H'$
\cr
R'^{-1}\Rightarrow {\rm orderly} & $\alpha^{1/2} v_H'>u$} ,
\end{equation}
where $u$ is the random velocity of the accreted bodies.
Whether growth is in the orderly, neutral, or runaway regime is
crucial for understanding how the mass spectrum evolves.  However,
it is less important for following the evolution in the two-groups
approximation (see Section \ref{sec:twogroups}), where big bodies are defined
as those that dominate the stirring and small bodies as those that
dominate the mass density.

Runaway growth was discussed in Safronov (1972, ch. 9, p. 105).  
However, it is usually attributed to later works by Greenberg et al. (1978) and Wetherill \& Stewart (1989), who
found that the mass spectra in their particle-in-a-box simulations 
developed high-mass tails containing a
small number of bodies.
Wetherill \& Stewart (1989, 1993)
 claimed that dynamical friction was essential for
runaway growth; this view is now widely accepted, although 
it is misleading.

Runaway growth requires $v_H^\prime<u<v_{\rm esc}^\prime$.
Dynamical
friction, which reduces $v^\prime$ relative to $u$, does not affect
these inequalities. 
Wetherill \& Stewart's conclusion was a consequence of 
the initial conditions
they chose for their simulations. 
Initial velocities of all bodies were set comparable to the escape speed from the biggest body.
But the biggest body was just outside of the runaway regime---its escape velocity 
was not sufficiently larger than its velocity relative to other bodies.
By reducing the random velocity of the biggest body, dynamical friction tipped
the balance in favor of runaway growth.

Adding the dependence on $u$ to Equation \ref{eq:runaway} in the
runaway regime results in
\begin{equation}
{1\over R'}{dR'\over dt}\propto {R'\over u^2} \ \ \ , \ v'_{\rm esc}>u>v'_H \ .
\label{eq:runaway2}
\end{equation}
If $u$ does not evolve, 
\begin{equation}
R'(t)={R^\prime(0)\over 1-t/t_\infty} \ ,
\end{equation}
for $t<t_\infty$, where $t_\infty$ depends on $R^\prime(0)$. However,
the divergence of $R^\prime$ in finite time does not typify runaway
growth. Typically, $u$ increases on the same timescale as $R$, the radius
of the big bodies that viscously stir. For example, in the
collisionless case, $u\propto \Sigma^{1/2}R$ (Equation \ref{eq:u}), and
for constant $\Sigma$, $u\propto R\propto t$ (see Equation \ref{eq:tR}).
Inserting $u\propto t$ into Equation \ref{eq:runaway2} shows that as $t\rightarrow\infty$
growth slows down, i.e., $(1/R')dR'/dt\propto 1/t^2$ decreases with increasing $t$.
However, the evolution is still in the
runaway regime because two big bodies of radii $R_1'$ and $R_2'$ grow
at the relative rates
\begin{equation}
{d\ln R_1'/dt\over d\ln R_2'/dt}= {R_1'\over R_2'} \ ,
\end{equation}
which implies that the radius of the larger body runs away from that
of the smaller. Runaway addresses how the radii of two large
bodies increase with respect to each other and not how each
individually evolves in time. In the literature, these two distinct behaviors are frequently
confused.  
For example, it is often stated that runaway ends when big bodies are more effective at
viscous stirring than are small ones (e.g., \citen{IM93}; \citen{Raf03c}; Thommes, Duncan \& Levison 2003). This leads to the condition $\Sigma M\sim\sigma m$ for when runaway ends.
But the evolution can still be in the runaway regime while big bodies dominate the stirring.
Although $\Sigma M\gtrsim \sigma m$ is a necessary condidtion for oligarchy to
occur (see below), it is not sufficient.

If runaway does not end when large bodies start stirring small ones, when does it end?
There are at least three possibilities:  ({\it a}) when the large bodies grow sufficiently such that the small bodies become sub-Hill (subsequent growth is neutral);  ({\it b}) when the small bodies' speed grows beyond
the escape speed from big bodies 
(subsequent growth is orderly);
({\it c}) when  oligarchy begins (discussed next).

\section{Dispersion-Dominated Oligarchy, $u>v_H$}
\label{sec:superhilloligarchy}

\subsection{Transition from Runaway to Oligarchy}

At late times, each big body is solely responsible for the viscous
stirring of all bodies that cross its orbit; each big body heats its
own food.  The random velocities of small bodies that
cross the orbit of a big body depend on the big body's radius $R$; that
is, $u=u(R)$.
Provided $u(R)\propto R^\gamma$ with $\gamma> 1/2$, the radii of big
bodies converge (see Equation \ref{eq:runaway2}). Below, we show that
$\gamma=6/5$,
assuming that the size of the small bodies
does not evolve
(see Equation \ref{eq:ucolthird}), although this is a questionable assumption.  Consequently, growth changes from runaway to
orderly. Kokubo \& Ida (1998) coined the term oligarchy to describe this
situation.  Oligarchy may be the final stage of planet formation
in the outer Solar System, although oligarchy with $u<v_H$ (see Section 10) may be the more relevant case.  The final
size of the oligarchs is insensitive to their earlier evolution.

\subsection{How Oligarchy Proceeds: Battling Oligarchs}

Each oligarch dominates an annulus of width $u/\Omega$ within
which the small bodies cross its orbit.  As the oligarchs 
increase $u$ by viscous stirring, 
 two neighboring annuli---each with its own
locally dominant big body---merge.  Once the two big bodies enter the
same feeding zone, they share small bodies, so the ratio of their
growth rates is equal to the ratio of their radii.  Then the radius of
the larger body runs away from that of the smaller one. 
Even though oligarchs in separate annuli experience orderly growth, those
that enter the same annulus experience runaway growth.
This behavior continues
on ever larger scales until nearly all the small bodies are consumed. Thus oligarchs maintain a separation of  $\Delta a\sim u/\Omega$,
and their surface density is
\begin{eqnarray}
\Sigma&\sim& {M\over a \Delta a} 
\sim
{\rho R^2\alpha\over   a}
{v_H\over u}
 \  ,    \ \ \ u>v_H .
\label{eq:Sigma}
\end{eqnarray}
This equation fixes one of the free parameters discussed 
in Section 7.  The
chief remaining free parameter is the size of the small bodies.

Our description of oligarchy is in accord with that of Ida \& Makino (1993).
By contrast, Kokubo \& Ida (1998) stated that, as the oligarchs grow, ``orbital
repulsion keeps their orbital separations wider than about 5 Hill
radii.'' They base this on N-body simulations containing a few
oligarchs embedded in a sea of small bodies that dominate the surface
density. Kokubo \& Ida (1998) attributed the increasing orbital separation
of the oligarchs to their mutual repulsion. However, this must result
because
their simulation took place
within a narrow annulus. 
Orbital repulsion just depends on the surface density distribution and 
is independent of the sizes of the bodies.
We do not expect neighboring
oligarchs to move apart as they grow---there is no place for them to go. 
Clearly, the number of oligarchs must decrease with time.
In  the version of oligarchy by Kokubo \& Ida (1998), neighboring oligarchs rule in harmony.  In our version,
oligarchs continually battle their neighbors for local dominance in a
winner-takes-all war.  So ``battling oligarchs'' might be a more
appropriate term to describe this phase of planet formation.
Indeed, competition between oligarchs has been observed in the numerical simulations
of Kokubo \& Ida (2000).

\subsection{When Oligarchy Ends: Isolation Radius and Isolation Time}
\label{sec:whenoligarchyends}

We refer to the epoch when nearly all the small bodies have been consumed
as isolation. 
At isolation, $\Sigma$ and $R$, respectively, are given by
 $\Sigma_{\rm iso}\sim \sigma$ and
(from Equation \ref{eq:Sigma})
\begin{equation}
R_{\rm iso}\sim 
\left(
{\sigma\over\rho}
{a\over\alpha}
{u\over v_H}
\right)^{1/2} \  .
\label{eq:rcompletiona}
\end{equation}
This occurs on the timescale (Equation \ref{eq:masslast})
\begin{equation}
t_{\rm iso}\sim \Omega^{-1}\alpha{\rho\over \sigma} 
\left({u\over v_H}
\right)^2
R_{\rm iso}
\end{equation}

We have yet to determine $u$ at isolation.
  Small-body velocities are
likely damped by inelastic collisions.\footnote{We neglect gas drag
here but show in Section \ref{sec:gasdrag} that it can be accounted for by
replacing $s$ with a smaller effective radius.}  Otherwise, Uranus and
Neptune would have taken far too long to form (see Section \ref{sec:smallbodies}).
Balancing viscous stirring by an oligarch with inelastic collisions
as done in Equation \ref{eq:ucolsecond}, we find $u/v_H\sim
\alpha^{-1/2}(s/R)^{1/4}$.  Combining this with Equation
\ref{eq:rcompletiona}, we can solve for
\begin{eqnarray}
{u\over v_H}\Big\vert_{\rm iso}  &\sim& 
\alpha^{-1/3} a^{-1/9}
\left(
{\sigma \over\rho}
\right)^{-1/9}
s^{2/9}
 \ , 
 \label{eq:oliu}
\\
\label{eq:olir}
R_{\rm iso}&\sim&
\alpha^{-2/3}a^{4/9}
\left(
{\sigma\over\rho}
\right)^{4/9}
s^{1/9} \ .
\end{eqnarray}
The isolation time is then
\begin{eqnarray}
t_{\rm iso}&\sim&
\Omega^{-1}\alpha^{-1/3}
a^{2/9}
\left(
{\sigma \over\rho}
\right)^{-7/9}
s^{5/9} \ .\label{eq:tcompletiona}
\end{eqnarray}

\subsubsection{Numerical Expressions with $s=1$ km}

In Appendix C (Section \ref{subsec:toomre}), we calculate the radius of the
largest body that can condense out of a gravitationally unstable cold
disk while conserving its internal angular momentum. In the minimum
mass solar nebula, this critical radius is $s\sim 1\km$, independent
of $a$.  Motivated by this result, we set $s=1\km$ in the above
formulae to obtain numerical expressions. 
Setting $\sigma=\sigma_{\rm MMSN}$ (Equation \ref{eq:mmsn})
then yields
\begin{eqnarray}
{u\over v_H}\Big\vert_{\rm iso}&\sim& 
 4
 \left({a\over 10\ {\rm
AU}}\right)^{7/18} 
\label{eq:ucompletion} \\
R_{\rm iso} &\sim&  10,
000  \left({a\over 10\ {\rm AU}}\right)^{4/9} \km
\label{eq:olir1a} \\
t_{\rm iso} &\sim&  50 \left({a\over 10\ {\rm
AU}}\right)^{26/9} \Myr \ .
\label{eq:olit}
\end{eqnarray}
With $a\sim 30$ AU, oligarchic growth can produce $16,000$-km
bodies in 1 Gyr out of the minimum mass solar nebula.  Nonetheless, we regard this mode of growth as implausible.  The small bodies' random
speed $u\sim $  km/s greatly exceeds their surface escape speed, $u_{\rm esc}\sim $ m/s.
Collisions at these great speeds are almost certainly destructive. Hence $s$ is likely much smaller than 1 km.  
Collisional fragmentation has been considered by
Inaba, Wetherill \& Ikoma (2003), 
Kenyon \& Bromley (2004) and Rafikov (2003e).

\subsubsection{Varying $s$ at 30 AU}

We can see how $R_{\rm iso}$ and $t_{\rm iso}$ vary with $s$
from Equations \ref{eq:olir}--\ref{eq:tcompletiona}.
But in deriving
these equations, we made a number of assumptions that restrict their range of
validity.
First, we assumed $u>v_H$ at isolation,
which implies $s>R\alpha^2$, i.e.,
\begin{equation}
s > 30 \cm \ .
\label{eq:slower}
\end{equation}
at $a=30$ AU.
Second, we assumed $u<\Omega a$; otherwise small bodies would
be ejected.  This
implies
\begin{equation}
s<200 \km \ .
\end{equation}
We also made the following assumptions, which always hold at
isolation at 30 AU as long as $30 \cm<s<200 \km$: ({\it a}) $u>u_{\rm
esc}$, ({\it b}) $u<v_{\rm esc}$, ({\it c}) the disk of small bodies is
gravitationally stable, and ({\it d}) the small-body disk is optically
thin. However, $t_{\rm iso}\lesssim 5$ Gyr
 sets the more
stringent upper limit of
 \begin{equation}
 s < 10 \km \ .
 \end{equation}
Because of collisional fragmentation, the upper limits on $s$ are largely irrelevant.
But if small bodies are ground down to sizes smaller than $\sim30$ cm, then 
oligarchy with $u<v_H$ must be considered (see Section \ref{sec:subhilloligarchy}).

\subsection{A Worked Example}
\label{sec:worked}

We now return to the early stages of oligarchic growth and follow the
subsequent evolution of $R$ and $u$.  
Here we fix $a=30$ AU, which gives
$\Omega^{-1}\sim 30$ y and $\sigma\sim 0.2$ g cm$^{-2}$ for the
MMSN (Equation \ref{eq:mmsn})
, we set $\Sigma$ to its
oligarchic expression (Equation \ref{eq:Sigma}), and 
we set
$s=1$ km in the early stages.  At later stages, fragmentation reduces $s$.

\subsubsection{Stage 1: Collisionless Oligarchy}
\label{subsubsec:collisionless}

The value of $R$ at which oligarchy starts depends on the runaway mass
spectrum for which we lack a simple model.  We consider the evolution after
$R$ reaches 100 km, assuming that oligarchy has begun by then.
 At this stage, viscous
stirring proceeds faster than collisional damping.  Ignoring
collisions, $u$ and $R$ both increase on the same timescale.  
From Equations \ref{eq:u} and \ref{eq:tR},
\begin{eqnarray}
{u\over v_H}&\sim& \left({R\over 100 \ {\rm km}}\right)^{2/3} \ ,
\label{eq:uoli}
\\ R&\sim&  200\left( {t\over {\rm
Myr}} \right)^{3/7}\km \ .
\end{eqnarray}
We chose $R > 100$ km to avoid
the sub-Hill equations here.

\subsubsection{Transition from Collisionless to Collisional Evolution} 
\label{sec:transition}

As collisionless oligarchy proceeds, $u$ increases at an
ever-decreasing rate: $(1/u)du/dt\propto 1/t$.  By contrast, the
collision rate is constant, so eventually collisional damping balances
viscous stirring.  This happens on the collision time (Equation \ref{eq:tcollide}):
\begin{equation}
t_{\rm transition}\sim t_{\rm collide}
\sim 10 \ {\rm Myr} \ ,
\end{equation}
when $R$ has reached
\begin{eqnarray}
R_{\rm transition}\sim 
500{\rm \ km} \ .
\label{eq:rtrans}
\end{eqnarray}
At this time, $u_{\rm transition}\sim 10$ m/s, which exceeds the surface escape speed
from kilometer-sized bodies, $u_{\rm esc}\sim 1$ m/s.  Hence collisions might be destructive as
soon as they commence.  But it is also possible that intermolecular forces provide some cohesion
to the small bodies and that the threshold speed for destruction is as high 
as 100 m/s.
For definiteness, we assume that collisions are not destructive at the transition
to  collisional oligarchy,
although this is highly uncertain.

\subsubsection{Stage 2: Collisional Oligarchy}
\label{sec:collisional}

Equating the rates of viscous stirring and collisional damping
 (Equation \ref{eq:ucolsecond}) yields
\begin{equation}
{u\over v_H}\sim 
2 \left({R\over 100\ {\rm km}}\right)^{1/5}
\ .
\label{eq:ucolthird}
\end{equation} 
and, from Equation \ref{eq:tcoll},
\begin{equation}
R\sim  10,000 \left( {t\over {\rm Gyr}}
\right)^{5/7}\km  \ .
\label{eq:r57}
\end{equation}
Collisional evolution continues until isolation
($\Sigma\sim\sigma$). If small bodies kept $s\sim$ 1 km, isolation would
occur when
$t\sim 1$ Gyr and $R\sim10,000$ km.  But by then, $u$ would be 
$\sim1$ km/s.
Collisions at these speeds would undoubtedly be destructive.

 \subsubsection{Stage 3: Collisional Oligarchy with Fragmented Bodies}

At some stage before the oligarchs become planets, small bodies fragment, reducing
$s$ far below 1 km.  Smaller $s$ implies smaller $u$ and, hence, faster accretion:
If we reintroduce the dependence of Equation \ref{eq:r57} on $s$ and solve for $t$, we find
$t\sim (R/10,000\km)^{7/5}(s/\km)^{2/5}$ Gyr.   If $s\sim30$
cm, planet-sized bodies form in 100 Myr.   In fact, it is possible that $s\ll 30$ cm, in which
case the planets form even faster.  But  then, $u<v_H$ at isolation (Equation \ref{eq:slower}).  
This leads us to examine oligarchy in the shear-dominated limit.

\section{Shear-Dominated Oligarchy: $u<v_H$}
\label{sec:subhilloligarchy}

Oligarchy with $u<v_H$ is not well treated in the literature.
It differs qualitatively from oligarchy with $u>v_H$ in several respects:
\begin{itemize}
\item As time progresses, oligarchs that previously were isolated begin
to compete with their neighbors to accrete the same small bodies.  For
$u<v_H$, growth is in the neutral regime, and the ratio
of the radii of big bodies maintains a constant value.
By contrast, for $u>v_H$, growth is in the runaway regime, and the ratio of radii diverges.
\item 
For $u<v_H$, an oligarch accretes from an annulus of width $\sim R_H$ and stirs a
wider annulus of width $\sim (v_H/u)^{1/2}R_H$.
By contrast, for
$u>v_H$, an oligarch both excites and accretes small bodies
from an annulus of width $\sim u\Omega^{-1}$.
\item For $u<v_H$, the oligarchs lie in a flat disk and 
may
accrete each other efficiently.
\end{itemize}

Each of these items raises a question about oligarchy in the shear
dominated regime: 
\begin{itemize}
\item Can shear-dominated oligarchy be maintained as oligarchs grow?
\item Is the spacing between oligarchs set by their feeding zones or by
their excitation zones?
\item  Do oligarchs grow more by coalescing than by eating small bodies?
\end{itemize}

In the shear-dominated limit, we
 show that oligarchy,
 defined as a single big body in its feeding zone 
 of width $\sim R_H$, is 
usually maintained, particularly at late times.
As each big body accretes small bodies, its feeding zone  expands.
 Once neighboring oligarchs enter each other's feeding zones, they quickly coalesce.
We can see this as follows: 
Oligarchs coalesce
 at the rate
$\alpha^{-3/2}\Sigma\Omega/\rho R$, and they
 grow by accreting  small bodies at the rate
$\alpha^{-1}(v_H/u)\sigma\Omega/\rho R$.
When neighboring oligarchs enter each others' feeding zones, growth
by coalescence is faster than that by accretion of small bodies, provided
\begin{equation}
{\Sigma \over \sigma} \alpha^{-1/2} {u\over v_H} >1 \ ,
\end{equation}
which holds for $\Sigma$ not much smaller than $\sigma$.\footnote{Except  if $u<\alpha^{1/2}v_H$, when both rates are comparable.  The small-body accretion rate saturates
for $u<\alpha^{1/2}v_H$.  This limit is only applicable when small bodies are centimeter sized or smaller.}
This ensures that there is typically only one big
body within each feeding zone and justifies our
definition.  
Growth proceeds in two phases:  Oligarchs slowly grow by accreting small bodies.  
Upon doubling their mass, they enter their neighbors' feeding zones.
Rapid coalescence follows.
On average,
growth by coalescence is comparable to that by 
accretion of small 
bodies.

When $u<v_H$,  an oligarch stirs a region 
of size $(v_H/u)^{1/2}R_H$, which is
larger than
its feeding zone. 
Within that stirring region, oligarchs keep their original size ratio. Outside that region, there is 
an effect similar to that in dispersion-dominated oligarchy,
where oligarchs grow to similar sizes because each oligarch regulates its own growth
by heating its own food. 
If stirring in one heating 
zone is large, growth in that zone slows down. Thus oligarchs within each 
heating
zone collectively regulate their own growth, and 
$\Sigma$ tends to become uniform on scales larger than
the stirring scale $(v_H/u)^{1/2}R_H$.

Because the oligarchs' separation is $\Delta a\sim R_H$, their surface density is
\begin{equation}
\Sigma\sim {M\over a\Delta a}\sim 
{\rho R^2\alpha\over a} \ \ , \ \ u<v_H \ \ .
\end{equation}
This is the standard expression used
in the literature, 
but it has been incorrectly applied to the
 case $u>v_H$
(\citen{KI98}; Thommes, Duncan \& Levison 2003\nocite{Thommes03}).
At the end of sub-Hill oligarchy, when $\Sigma\sim\sigma$, the final value
of the oligarchs' radii is given by equating the above expression with $\sigma$, 
which gives the isolation radius 
\begin{eqnarray}
R_{\rm iso} \sim\left( {\sigma
a\over\rho\alpha}\right)^{1/2} &\sim &
12,000\left(
a\over 10{\rm\ AU} \right)^{1/4}\km \ .
\label{eq:iso2}
\end{eqnarray}
In the numerical expression, we have included factors of order unity (Equation \ref{eq:isolationradius}).
At isolation,  $u<v_H$ only if $s<200(a/{\rm 10\ AU})^{-7/4}\cm$.
 Larger values of $s$ are compatible with $u<v_H$
at earlier stages of planet growth.

\subsection{The Fastest Oligarchy}

The fastest possible accretion rate is faster than the geometrical 
rate by 
$\alpha^{-3/2}$,
and occurs for $u\sim\alpha^{1/2}v_H$.  Therefore
\begin{equation}
\label{eq:tfastest}
t_{\rm fastest}\sim
\alpha^{3/2} {\rho R \over \sigma \Omega} 
\sim
30,000\left(
{a\over {\ 10\rm AU}}
\right)^{7/4}\yr.
\end{equation}
More precisely, this
is the shortest time 
during which an oligarch could double in mass.
Oligarchy can proceed so fast that
 more time might be spent in forming kilometer-sized small bodies than  in growing these
bodies into planets.

Gravitational stability of the disk (see Section \ref{subsec:toomre}) dictates a minimum 
velocity dispersion of 
\begin{equation}
u_{\rm min}\sim
{\sigma\Omega\over\alpha^{3}\rho} 
\sim 1{\rm m/s}.
\end{equation}
With our nominal parameters, this is approximately equal to $\alpha^{1/2}v_H$ for bodies with the isolation radius (Equation \ref{eq:iso2}).
 It is not clear whether the short timescale
of Equation \ref{eq:tfastest} can materialize. 
If $u\sim u_{\min}>\alpha^{1/2}v_H$, the oligarchs grow at a constant rate (i.e.
exponentially) on the timescale
\begin{equation}
t_{\rm fastest,\ stability\ limited} \sim \alpha^{-1} \Omega^{-1}
\sim 10,000 \left( a\over 10\au\right)^2 \yr. 
\label{eq:tstability}
\end{equation}
To keep $u\sim u_{\rm min}$  with collisions, $s$ must be
\begin{equation}
s \sim {\sigma\over\rho}\sim \left( {a\over 10{\ \rm
AU}} \right)^{-3/2}\cm 
\label{eq:sstability}
\end{equation}
when $\Sigma\sim\sigma$.

Rafikov (2003e) showed that accretion is very fast if the accreted bodies are very small
and their random velocities are damped by gas drag.
Lissauer's (1987) expression for the fastest growth rate
 differs from ours. He assumed 
that the lowest $u$ possible is $v_H$ and, hence, that the fastest focusing factor is $\alpha^{-1}$.
He argued that in a cold disk every encounter with a large body results in a velocity 
of $v_H$, so $u\gtrsim v_H$.  But this is incorrect because collisional damping between encounters
can maintain $u<v_H$.

\section{Beyond Oligarchy}
\label{s:after}
\label{ss:endoligarchy}

Relatively little attention has been paid to stages of planet
formation beyond oligarchy.  Chambers (2001) reported results from an
N-body simulation initialized with several-dozen bodies separated by a
few times their Hill radii on coplanar, circular orbits. This setup is
unstable; eccentricities and inclinations quickly develop. This raises
two important questions: When in the course of oligarchic growth does
this instability occur? And what is its outcome?  These questions and
others are addressed by Goldreich, Lithwick \& Sari (2004), which the 
discussion here
closely follows.
We start by showing that oligarchy ends, and instability kicks in,
when $\Sigma\approx \sigma$. This result applies to accretion in both
shear- and dispersion-dominated regimes.

\subsection{Before Oligarchy Ends: $\Sigma<\sigma$}

When $\Sigma<\sigma$,
the oligarchs' velocity is set by a balance between mutual viscous
stirring and dynamical friction with small bodies.  
If $u>v_H$, this results in 
\begin{equation}
v\sim u\left(
{\Sigma\over\sigma}
\right)^{1/4}  \ \ , \ \ u>v_H
\end{equation}
(see Equation \ref{eq:vitou}).
If $u<v_H$,  then, as we shall see below, $v<v_H$ as well.
So viscous stirring excites $v$ at the rate
$\alpha^{-2}(\Sigma\Omega/\rho R)v_H/v$ 
(Equation \ref{eq:Fvs}), and dynamical friction damps
it at the rate
 $\alpha^{-2}\sigma\Omega/\rho R$ (Section \ref{s:v>u}). 
Equating stirring and damping yields
\begin{equation}
v\sim v_H {\Sigma \over \sigma} \,  \ \ , \ \ u<v_H \ \ .
\label{eq:vnusd}
\end{equation}
This justifies the use of rates appropriate to $v<v_H$.

\subsection{Instability Of Protoplanet's Velocity Dispersion: $\Sigma>\sigma$}
\label{subsec:instab}

As soon as $\Sigma>\sigma$, the velocity dispersion of the big bodies
destabilizes.  This occurs because the typical relative velocity
between a big and small body is $v>u$, so the velocity evolution
equation becomes (see Section \ref{s:v>u}):
\begin{equation}
{1\over v}{dv\over dt}= {\left(\Sigma -\sigma \right)\Omega\over
\rho R}\alpha^{-2} \left( {v_H\over v} \right)^4 , 
\label{eq:vinstability}
\end{equation}
regardless of whether $u<v_H$ or $u>v_H$.
Thus when $\Sigma >\sigma$, big bodies are heated faster than they are
cooled.  This marks the end of oligarchy. As $v$ increases, heating
and cooling both slow down, but heating always dominates cooling.
Rather quickly, the orbits of neighboring big bodies cross.

Because it is based on approximate rates for viscous stirring and
dynamical friction, the criterion, $\Sigma\sim \sigma$, for the onset
of velocity instability is also approximate.  N-body simulations of
oligarchy with the addition of accurate analytic expressions for
dynamical friction are needed to identify the critical surface density
ratio.

The consequence of the instability in the velocity dispersion differs
according to which is larger, the escape velocity from the surfaces of
the planets that ultimately form or the escape velocity from their
orbits. The ratio of these two escape velocities, $v_{\rm esc}/\Omega a$,  is
\begin{equation}
{\cal R}\sim \cases{0.3 & for $a=1\au\,\, \& \,\, M_p=M_\oplus$ \cr
2.3 & for $a=25\au\,\, \& \,\, M_p=15 M_\oplus$}\, ,
\label{eq:escratio}
\end{equation}
where $M_p$ is the mass of a planet and $M_\oplus$ is Earth's mass.

\subsubsection{Inner Solar System, ${\cal R}\ll 1$: Coalescence}

In regions where ${\cal R}\ll 1$, the big bodies' velocity dispersion
increases until it becomes comparable to the escape velocity from their
surfaces. At this point, they begin to collide and
coalesce. Coalescence slows as the number of big bodies decreases and
their individual masses increase.

The timescale for the formation of planet-sized bodies with radius
$R_{\rm p}$ whose orbits are separated by $\sim a$ is
\begin{equation}
t_{\rm coag}\sim \left(\rho R_p\over \sigma \Omega\right)\sim 10^8\yr
\,\, {\rm at\,\,} a=1\au\,\, {\rm for} \,\, R_p=R_\oplus\, .
\label{eq:tcoag}
\end{equation}
At a separation of order $a$, mutual interactions no longer produce
chaotic perturbations. Indeed, the N-body simulations of
Chambers (2001) produce stable systems on a timescale
similar to $t_{\rm coag}$.

What happens to the small bodies while the big ones are colliding and
coalescing? On one hand, a significant fraction collide with and are
accreted by
big bodies.  On the other hand, new small bodies are created in grazing
collisions
between big ones, ensuring that a significant
residual
population of small bodies persists until the end of coalescence.

\subsubsection{Outer Solar System, ${\cal R}\gg 1$: Ejection}

In regions where ${\cal R}\gg 1$, $v$ reaches the orbital speed
$\Omega a$.  Some fraction of the big bodies becomes detached from the
planetary system and either takes up residence in the Oort cloud or
escapes from the Sun. This continues until mutual interactions among
the surviving big bodies are no longer capable of driving large-scale
chaos.

We estimate the ejection timescale as
\begin{equation}
t_{\rm eject}\sim {0.1\over\Omega}\left(M_\odot\over
M_p\right)^2\sim 10^{9}
{\yr}\,\, {\rm at\,\,} a=25\au \, .
\label{eq:tesc}
\end{equation}
Shoemaker \& Wolfe (1984) and Dones et al. (2004) reported similar timescales for the
ejection of test particles placed on orbits between Uranus and
Neptune, the former authors from a crude impulsive treatment of scattering and
the latter authors from N-body integrations. A shorter timescale might apply
if bodies were transferred to and then ejected by Jupiter and
Saturn.

As the random velocity of a big body increases, the rate at which it
accretes small bodies declines. Thus a substantial surface density of
small bodies is likely to remain after most the big bodies have
been ejected.

\subsection{Orbit Regularization}
\label{ss:regularization}

Either coagulation or ejection is likely to end with the surviving big
bodies moving on orbits with eccentricities and inclinations of order
${\cal R}\sim 0.3$ in the inner-planet system and of order unity in
the outer-planet system. Chambers (2001) saw the former in N-body simulations.  Such orbits do not resemble those of planets in the Solar System.  In reality, dynamical friction by the residual small bodies
tends to circularize and flatten the orbits of the surviving
protoplanets. 

As the surviving planets cool down, their orbits no longer cross. Gaps
may open in the disk of small bodies around the planets' orbits.
Dynamical friction continues to act after gap opening. 
Angular momentum and energy are transferred between the planet and the
disk of small particles by torques that the planet exerts at
Lindblad and corotation resonances. Ward  \& Hahn (1998, 2003) used the standard
torque formula (Goldreich \& Tremaine 1980) and concluded that the most potent
contributions to the damping of eccentricity and inclination are due
to torques at apsidal and nodal resonances. 
The analysis of Goldreich \& Sari (2003), when applied to the cold planetesimal
disks, suggests that the apsidal torque will be the most potent one,
albeit with lower magnitude than Ward \& Hahn (1998, 2003) had calculated.

\subsection{Clean Up}

What was the fate of the residual small bodies that remained after the
protoplanets had settled onto stable orbits?  Velocity instability
started when small bodies and protoplanets contributed comparably to
the overall surface density.  But today, the mass in small bodies is
much less than that in planets.
Clean up was both the last and longest stage in the evolution of the Solar System. It is ongoing in both the asteroid and Kuiper belts. The
Oort comet cloud was probably populated during this stage.  Some
thoughts on how clean up was achieved are sketched in
Goldreich, Lithwick \& Sari (2004).

\section{Summary}

We begin this review (Sections \ref{sec:thehillsphere}--\ref{sec:massgrowthrate}) with an overview of the physics of coagulation: how bodies in a circumsolar disk grow by accreting each other, and how they stir and damp each other's velocities through viscous stirring and dynamical friction.  We present order-of-magnitude derivations for relevant
formulae that capture the underlying physics.
For $u<v_H$, a small
fraction of the bodies contains a large fraction of the energy in
random motions, and inclinations are stirred less than eccentricities, which can lead to a very flat disk. In this limit, viscous stirring has not been correctly
treated in the literature.

In Sections 6--10, we focus
on the growth of Uranus and Neptune, whose masses and formation times greatly constrain how they formed.   Because these planets do not contain a lot of gas, we neglect gas drag.
But in Appendix A, we show that gas drag can simply be accounted for by setting the effective size of small bodies below its true value.  

A simple argument (see Section \ref{sec:growthofplanets}) shows that Uranus and Neptune grew by accreting bodies that were smaller than 1 km.  Had the accreted bodies been bigger than 1 km, they 
would not have collided with each other sufficiently frequently to damp their speeds, which are
vigorously stirred by the planetary embryos.  Accretion of such
hot bodies would have taken longer than the age of the Solar System.

In Section \ref{sec:solving}, we consider accretion more quantitatively.  The two-groups approximation 
is a valuable tool.  Often, only two groups of bodies---those with most of the mass and those that dominate viscous stirring---control the evolution of all others.  It is nearly trivial to write down the equations describing the evolution of the two groups and not difficult to solve them.  Because the rest of the bodies behave passively, their velocity spectrum is also easy to work out 
(see Appendix B).  The two-groups approximation is helpful in analyzing results from N-body simulations and particle-in-a-box simulations.  
However, we have been unable to arrive at a simple quantitative picture for how the
mass spectrum evolves.  Simulations are still needed for that.
In Section \ref{sec:orderlyneutral}, we address qualitatively the different ways in which the mass spectrum can evolve, by orderly, neutral, or runaway growth.

In Sections \ref{sec:superhilloligarchy}--\ref{sec:subhilloligarchy}, we discuss oligarchy with 
$u>v_H$ and with $u<v_H$.
Dispersion-dominated oligarchy begins when each big body heats its own food, i.e., when the
large bodies' orbits are separated by $u/\Omega$.  Because oligarchs regulate their own growth, their masses equalize.  We derive expressions for the final sizes of the oligarchs as well as their formation times.
The formation time is quite sensitive to the assumed value of the small bodies' radii.
Shear-dominated oligarchy
begins when the large bodies' orbits are separated by a few times their Hill sphere. 
Discussions of oligarchy in the literature mix elements  from dispersion-dominated and shear-dominated oligarchy. 
 We show that oligarchy pertains in both regimes,
 albeit with different behaviors. Oligarchy does not occur for very flat disks with $u<\alpha^{1/2}v_H$.

We outline two scenarios for the formation of Uranus and Neptune. If the accreted bodies were $\sim$1 km, then Uranus and Neptune would have formed in a few billion years (Equations  \ref{eq:tforma}, \ref{eq:olir1a},  and \ref{eq:olit}). 
Kilometer-sized bodies are a natural outcome of gravitational instability of a planetesimal disk (see Appendix C).
However, we do not view the solution with
kilometer-sized bodies as physical.  
Uranus and Neptune would excite the velocity dispersion of these bodies to such an extent that
they would shatter when they collided, thereby creating much smaller bodies.
This leads us to favor a scenario in 
which the accreted bodies were  much smaller than 1 km.  In the limit in which collisions reduced the radii of
the accreted bodies to
a few centimeters or less, the final
doubling of the masses of Uranus and Neptune could have taken 
less than one million years 
 (Equations  \ref{eq:tfastest}, \ref{eq:tstability}, and \ref{eq:sstability}) .
Although this resolves the timescale problem, it comes with its own caveat.
In the minimum mass solar nebula, shear-dominated accretion leads
to the formation of 10 small planets beyond Saturn (Equations \ref{eq:isolationradius} and  \ref{eq:iso2}).   Why, then, are we left with 2 large ones? We suggest that Uranus and
Neptune have collected most of their mass by the end of oligarchy. This requires an initial surface density of
a few times that of the MMSN.
The uncertainties are ({\it a}) the exact epoch when oligarchy ends
and  velocity instability begins, ({\it b}) the radial range from which oligarchs can accrete, and 
({\it c}) the amount of mass that surviving planets can accrete during the cleanup stage. Other planet-sized
objects probably formed in and were ejected from the outer Solar System after oligarchy.

We conclude with a discussion of post-oligarchic
 stages. Velocity
instability leads to ejection of bodies from the outer-planet
system and to their coagulation in the inner one. It ceases once the
number of protoplanets is sufficiently reduced. After that, dynamical
friction from the residual small bodies damps the protoplanets'
orbital eccentricities and inclinations.  The last and longest stage
is also the least understood one: cleanup of the remaining small
bodies. Wide gaps that form around the orbits of the surviving planets
inhibit accretion. Collisions prevent the outer planets from ejecting
small bodies from the outer-planet region. Very little attention has
been paid to cleanup. It is a frontier in planet formation worth
exploring.

\section{Appendix A: Neglected Effects}

\subsection{Gas Drag}
\label{sec:gasdrag}

One of the main effects of gas drag is to damp the random
velocities of small bodies.
This is simply accounted for in our formalism.
Damping by inelastic collisions is inversely proportional to $s$,
 and  $s$ appears only in 
 connection with
damping by inelastic collisions. 
To account for gas drag, it suffices to set $s$
to a value
smaller than the true size of the small bodies.  
As we have emphasized, the size of the
small bodies is highly uncertain.  
In the following, we estimate the gas-damping rate (for more complete treatments, see Adachi, Hayashi \& Nakazawa 1976,
Rafikov 2003e).

Gas drag damps a small body's random velocity in the time that
the body encounters a mass of gas equal to its own mass.  Thus
\begin{equation}
{ {\rm gas \ damping\  rate}\over {\rm collisional\  damping\  rate}
}\sim{\rho_{\rm gas}\over\rho_{\rm bodies}}{\Delta u_{\rm gas}\over u}\sim
{\sigma_{\rm gas}\over\sigma}
{\Delta u_{\rm gas}\over c_s} \ ,
\label{eq:gasdrag}
\end{equation}
where $\rho_{\rm gas}$ and $\rho_{\rm bodies}$ are the midplane mass
densities of the gas and the small bodies, $\sigma_{\rm gas}$ and $c_s$ are
the surface density and sound speed of the gas, and $\Delta
u_{\rm gas}$ is the speed at which a small body encounters the gas.
We assume that $u<c_s$, so the scale height of the gas disk is larger
than that of the small bodies. In fact, we assume below the slightly
more restrictive condition $u<c_s^2/\Omega a$, which
almost certainly holds within the first 10 Myr in the outer Solar
System.

We have yet to calculate $\Delta u_{\rm gas}$.  There are two limiting
regimes, depending on the size of the small body relative to the
mean free path of a gas molecule $l_{\rm mfp}$.
When $s<l_{\rm mfp}$, gas particles act as tiny bodies whose
random speed is $c_s$ (i.e., the free molecular drag, or Epstein,
regime), so $\Delta u_{\rm gas}\sim c_s$ and Equation
\ref{eq:gasdrag} becomes
\begin{equation}
{ {\rm gas \ damping\  rate}\over {\rm collisional\  damping\  rate}
}\sim {\sigma_{\rm gas}\over\sigma}\ \ , \ \  \ \ s<l_{\rm mfp} \ .
\label{eq:dragEpstein}
\end{equation}

For sufficiently large $s$, the gas can be treated as a collisional
fluid, and $\Delta u_{\rm gas}$ is the speed at which a small body
moves through the gas (the turbulent drag regime). There is an intermediate regime in which Stokes
(viscous) drag applies. We ignore this and consider the limiting case
of turbulent drag. Small bodies move relative to the gas because the
orbital speed of the gas around the Sun is sub-Keplerian owing to
thermal pressure support, i.e., $V_{\rm gas}^2\sim (\Omega
a)^2-c_s^2$, where $V_{\rm gas}$ is the gas's orbital speed. Thus
$\Delta u_{\rm gas}\sim \Omega a- V_{\rm gas}\sim
c_s^2/(\Omega a) $, to lowest order in $c_s/(\Omega
a)$.\footnote{As mentioned previously, we assume
$u<c_s^2/(\Omega a)$; otherwise, $\Delta u_{\rm gas}\sim
u$.}  Equation \ref{eq:gasdrag} becomes
\begin{equation}
{ {\rm gas \ damping\  rate}\over {\rm collisional\  damping\  rate}
}\sim 
{\sigma_{\rm gas}\over \sigma}
{c_s\over\Omega a}\ \ , \ \  \ \ s>
{\Omega a\over c_s }
l_{\rm mfp} \ ,
\label{eq:dragturb}
\end{equation}
where the limit on $s$ ensures that the Reynolds number exceeds unity.

Numerically,
\begin{equation}
{\sigma_{\rm gas}\over\sigma}\sim 100 \ ,
\end{equation}
for cosmic composition and condensates expected in the outer Solar
System.  We approximate the gas temperature by equating it to the local equilibrium
temperature, $T_{\rm gas}\sim T_\odot\alpha^{1/2}$, where
$T_\odot\sim 6,000\,$K  is the surface temperature of the Sun (for more-realistic models, see Chiang \& Goldreich 1997). This yields
$c_s\sim (k_B T_{\rm gas}/m_p)^{1/2} \sim 6
\alpha^{1/4}\km\s^{-1}$, where $k_B$ is Boltzmann's constant and
$m_p$ is the proton mass. Hence
\begin{equation}
{c_s\over\Omega a}\sim 0.1 
\left(
{a\over 10\ {\rm AU}}
\right)^{1/4} \ ,
\end{equation}
and
\begin{equation}
l_{\rm mfp}\sim {m_p \over\rho_{\rm gas}\sigma_{\rm x-section}}\sim 
300\   \left( {a\over 10{\rm\ AU}} \right)^{11/4} \  \cm,
\end{equation}
where $\sigma_{\rm x-section}\sim 10^{-15}\cm^2$ is the collisional
cross section of gas particles. Equation \ref{eq:dragEpstein} shows
that gas drag might damp the random velocities of small bodies as
much as two orders of magnitude faster than inelastic collisions.
Although the enhancement factors are not very sensitive to $a$,
the sizes of the bodies at which transitions between drag
regimes occur are
extremely
sensitive. Gas drag might also be important for the orbital decay of planetesimals
(Adachi, Hayashi \& Nakazawa 1976\nocite{AHN76}; \citen{Weiden77b}).  
In addition, if a protoplanet has a gaseous atmosphere, then gas drag on planetesimals that pass through its atmosphere can enhance the accretion rate (Inaba \& Ikoma 2003).  

A potentially more significant effect of gas drag is its role
in damping the random velocities of oligarchs. 
The velocities of large bodies can be damped by gas in the same
way that it is damped by small bodies.
Ward (1993)
showed that the rate of gas damping can be obtained from the damping rate due to small bodies by
substituting the surface density of gas for that of the
small bodies and the sound speed of the gas, $c_s$, for the
random velocity of the small bodies, $u$. The dispersion-dominated
region is relevant because at isolation $v_H<c_s<v_{\rm esc}$.
Moreover, we assume that $v<c_s$. By damping the oligarchs'
random velocities, gas drag could delay the onset of velocity instability
and enable the oligarchs to consume all the small bodies.

In the inner Solar System, it is possible---although highly
uncertain---that much of the gas survived until isolation. Then the
full velocity instability of the oligarchs would have been delayed
until the surface density of the gas declined to match that contributed by the
oligarchs. After that, the oligarchs would have excited their random
velocities up to their escape speeds.  Although most the small
bodies would have been accreted before this happened, plenty of new
ones created in glancing collisions could have damped the orbital
eccentricities and inclinations of the planets that finally
formed. Gas is unlikely to have survived long enough to contribute to the regularization of their orbits.

Uranus and Neptune
are believed to have
collected only a few Earth masses of nebular gas. So it is likely that
most the gas had disappeared prior to isolation in the outer Solar
System.

\subsection{Accretion of Gas}

How did the giant planets accrete their gas?
In the core accretion model, the cores of the gas giants first formed via coagulation. 
Once the cores were sufficiently massive, they gravitationally captured gas from the surrounding disk 
(Mizuno, Nakazawa \& Hayashi 1978; Pollack et al. 1996).\footnote{
Pollack et al. (1996) simulated portions of this process numerically.  But during growth by coagulation, 
they arbitrarily set the the random speed of the accreted material to its own escape speed; this is 
an underestimate because it neglects viscous stirring by the planetary embryo.  Hence, given the
sizes of the planetesimals that they assume (1 km and 100 km), they underestimate the time spent in the coagulation phase, which they say is $\sim$1--20 Myr for Jupiter, Saturn, and Uranus. Nonetheless, such fast accretion can occur if the accreted bodies are much smaller than 1 km.
}

Other formation
mechanisms have been proposed. Most notable is gravitational instability of a gas disk, without the prior formation of a core.
A sufficiently cold gas disk is gravitationally unstable 
and can
 collapse directly
into giant gaseous planets (Boss 1997, Mayer et al. 2002).\footnote{This is similar to the gravitational instability
of condensates (see Appendix C).}
Jupiter and Saturn could have formed in this way, but only if the
surface density was greater than the MMSN value by more than a factor of 10 \cite{GuillotGladman2000}.
Some extra-solar planets might also have formed in this manner.
Uranus and Neptune have only 10--20\% gas, so it is likely that they formed via coagulation.  

\subsection{Other Neglected Effects}

Jupiter and Saturn probably formed before Uranus and Neptune.  Therefore they viscously
stirred the planetesimals that were accreted onto Uranus and Neptune 
(\citen{KW00}; Kuchner, Brown \& Holman 2002\nocite{KBH02}).
We neglect this because, in the late stages of accretion, stirring by the embryos of Uranus and Neptune is more important.
We do not consider gap formation by a growing embryo \cite{Ward97,Raf01}, although gaps might hinder growth.
We also neglect the effects of solar radiation, such as radiation pressure, Poynting-Robertson drag, and the Yarkovsky effect (Burns, Lamy \& Soter 1979\nocite{BLS79}).

\section{Appendix B: Velocity Spectrum}
\label{sec:velocityspectrum}

\begin{figure}
\centerline{\epsfxsize=5.5in\epsffile{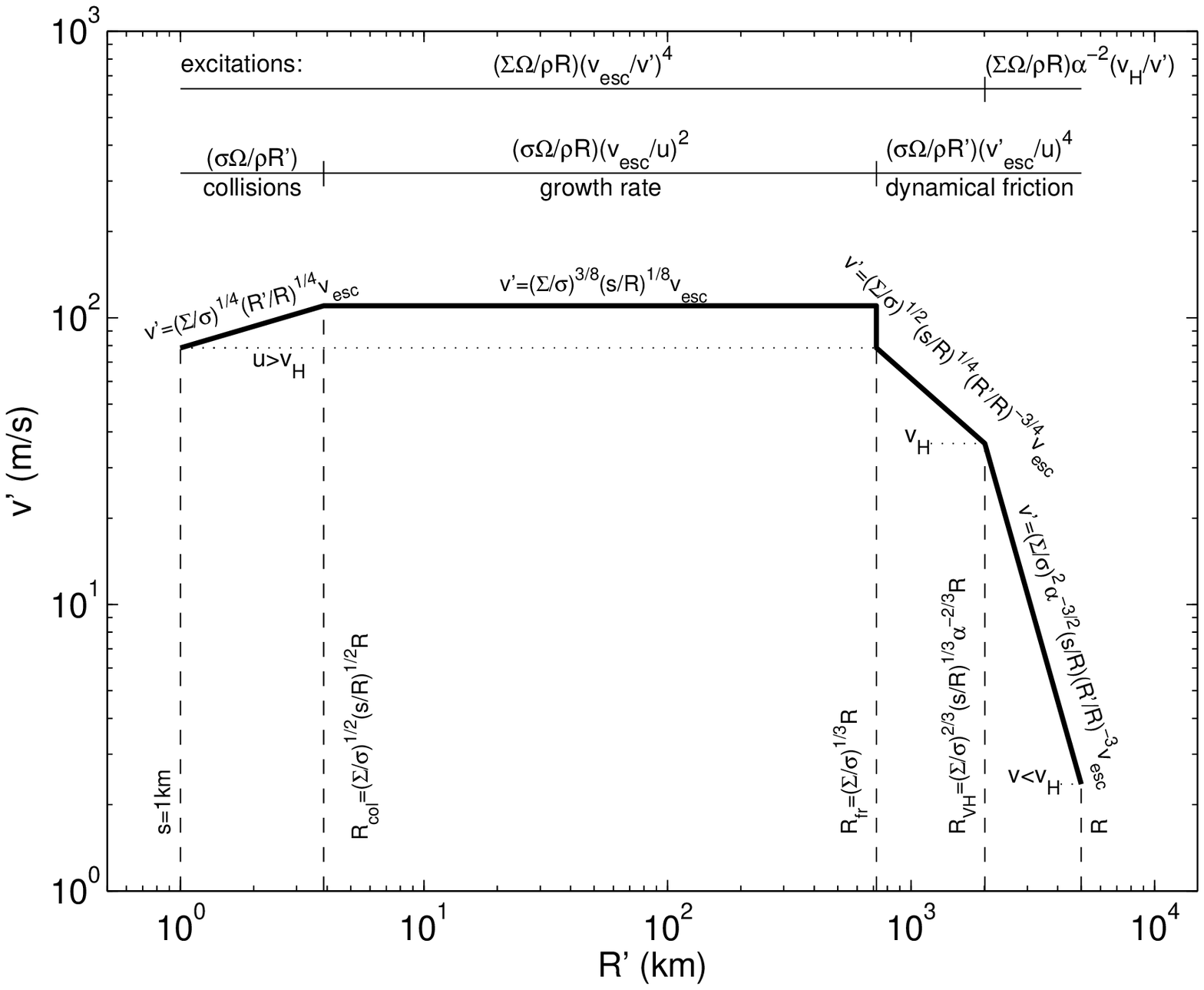}}
\caption{
Schematic plot of the random velocity spectrum, including
collisions. Small bodies are in the dispersion-dominated regime, and
large ones are in the shear-dominated regime. The spectrum comprises four power-law segments ({\it bold solid lines}). Rates at which
viscous stirring by big bodies excites random velocities are given
above the upper horizontal line. Rates that limit random velocities
are written above the lower horizontal line and identified immediately
below it. Each power-law segement of the spectrum is determined by
equating the excitation rate to the relevant limiting rate. Radii
bounding each segment are written along vertical dashed lines. Dotted,
horizontal line segments mark the values of the random velocities of
the small bodies, $u$, the random velocities of the large bodies, $v$,
and the Hill velocity of the large bodies, $v_H$. Starting from the
left, the first, second, and third segments show viscous stirring
in the dispersion-dominated regime limited by collisions, the growth
time of big bodies, and dynamical friction, respectively. The fourth
segment shows viscous stirring in the shear-dominated regime
limited by dynamical friction. Here we are plotting the typical random
velocity, which is smaller than the rms random velocity (described
in Section \ref{subsec:hillvelocity}). The discontinuity between the second
and third segments is closely related to the velocity instability that
occurs for $v>u$ (see Section  \ref{subsec:instab}).  Parameters chosen for
this figure are $R\sim 5,000\km$, $\sigma=0.2\gm\cm^{-2}$,
$\Sigma=0.003\sigma$, $a=30\au$, $\rho=1\gm\cm^{-3}$, and $s=1\km$.
}
\label{fig:spec2}
\end{figure}

We apply the two-groups approximation to determine the run of the
typical random velocity with radius $R^\prime$, where
$s<R^\prime<R$. Intermediate-sized
bodies behave passively. They merely respond to viscous stirring by
big bodies and dynamical friction by small ones.
This is similar to our discussion in Section 7.3.3, except that here
we consider big bodies with $v<v_H$ and small bodies that are collisional but
maintain $u>v_H$. This introduces two additional transitions: one
between collisional and collisionless behavior at $R_{col}$ and
another at $R_{v_H}$, where $v^\prime=v_H$. We find
\begin{equation}
R_{col}=\left( \Sigma \over \sigma \right)^{1/2}\left( s\over R\right)^{1/2}R \, ,
\end{equation}
\begin{equation}
R_{fr}=\left( \Sigma \over \sigma \right)^{1/3}R\,  ,
\end{equation} 
\begin{equation}
R_{v_H}=\left( \Sigma \over \sigma \right)^{2/3}\left( s\over
R\right)^{1/3}\alpha^{-2/3}R\, .
\end{equation}
Figure \ref{fig:spec2} displays the entire random velocity spectrum
for a specific choice of parameters: $\sigma=0.2\gm\cm^{-2}$,
$\Sigma=0.003\sigma$, $a=30\au$, $s=1\km$, and $R=5000\km$. The choice
of $R$ determines the time.

As discussed in Section \ref{subsec:hillvelocity} ,
the typical random velocity is similar to the velocity dispersion
except for $v^\prime<v_H$.
In the sub-Hill regime, $R_{v_H}<R'<R$, the typical random
velocity, $v'/v_H\sim (R'/R_{v_H})^{-3}$, is smaller than the rms
velocity $v^\prime_{\rm rms}/v_H\sim (R'/R_{v_H})^{-3/2}$. The latter
implies energy equipartition among bodies in the sub-Hill regime as
found in some simulations.  This equipartition has a nonstandard
origin. It results from a balance between viscous stirring by large
bodies and dynamical friction from small ones, where the random energy
of a big body is much larger
 than that of a small one  (see Rafikov 2003d).
 In the sub-Hill regime (see Section 
\ref{subsec:hillvelocity}), the distribution
function of random velocities at fixed $R^\prime$ between the typical
velocity and $v_H$ is a power law.

The spectral shape of the solution displayed in Figure \ref{fig:spec2}
requires $s<R_{col}<R_{fr}<R_{v_H}<R$, supplemented by $u>v_H$. It
also assumes that big bodies grow by accreting small ones, which holds
if
\begin{equation}
\left( v_H \over u \right)^2 > {\Sigma \over \sigma} \alpha^{-1/2}\, .
\label{eq:accretesmall}
\end{equation}
Equation \ref{eq:accretesmall} is marginally satisfied for the
parameter values chosen in Figure \ref{fig:spec2}. Other
consistency constraints are that big bodies dominate the stirring and
that small ones provide the dynamical friction.  It is not difficult
to choose size distributions that satisfy these constraints.


\section{Appendix C: Considerations Relevant to the Size of Small Bodies}
\label{sec:smallbodysize}

Without a compelling reason to favor any particular radius for the
small bodies, we have been treating $s$ as a free parameter.  A
smaller $s$ enhances the damping of $u$ by collisions and gas drag,
which speeds up accretion. We suggest that as the big bodies grow, the
small bodies undergo destructive collisions, which results in a
fragmentation cascade. However, there is a floor on $u$ below which
gravitational instabilities would lead to the rapid formation kilometer-sized bodies. 

\subsection{Destructive Collisions}
\label{subsec:destructive}

Fragmentation has been considered by, e.g., Inaba, Wetherill \& Ikoma (2003), Kenyon \& Bromley (2004), and Rafikov (2003e).
Even in the shear-dominated limit,
$u<v_H$, a small body that passes within $R_H$ of a big body has its
random velocity excited to order $v_H$,\footnote{The
typical value of $u$ exceeds $v_H$ during most of oligarchy unless $s$
is  smaller than 30 cm.} where
\begin{equation}
v_H\sim \Omega R_H\sim 10^2\left(R\over 10^2\km\right)\left(a\over
10\au\right)^{-1/2}\cm\s^{-1}\, .
\end{equation}
By comparison, the escape
speed from the surface of a small body is
\begin{equation}
u_{\rm esc}\sim (G\rho)^{1/2}s\sim 
10^2\left({s\over {\rm km}}\right)\cm\s^{-1}\, . 
\end{equation}
This comparison reveals that the random velocity induced by large protoplanets may suffice to break up kilometer-sized planetesimals. At what
stage this might happen depends on two unknowns: the size of
planetesimals at the start of runaway growth and their internal
strength \cite{Doh69}.

Suppose that at some stage the initial planetesimals begin to fragment.
As $s$ decreases, the collision rate 
$\sigma\Omega/(\rho s)$ (Equation \ref{eq:ucolfirst}) 
increases, which leads to an accelerating collisional cascade.
Although collisions damp $u$, the planetary embryo viscously stirs it.
Assuming that $u$ is set by a balance between collisional damping and viscous stirring,
$u/u_{\rm esc}$ increases as $s$ decreases.
Thus the collisions remain destructive so long as the stresses
they induce exceed the yield stress of the material (ice) composing the small bodies. Whether this occurs
before $u\sim u_{\rm min}$ (see Section \ref{subsec:toomre}) is unknown.

\subsection{Gravitational Instability in a Cold Keplerian Particle Disk}
\label{subsec:toomre}

In a disk of particles that orbit the Sun, if the particles' velocity dispersion $u$ falls below
the critical value
\begin{equation}
u_{\rm min}\sim {G\sigma\over\Omega}\sim 1 \m/\s \ , 
\label{eq:uminfirst}
\end{equation}
then the disk is gravitationally unstable \cite{BT87}.  
For the minimum mass solar nebula (Equation \ref{eq:mmsn}), which we used in the above numerical
expression, $u_{\rm min}$ is independent of $a$. 

We consider the evolution of a cold disk with $u=0$.
This disk is unstable.
Overdense
perturbations of size
\begin{equation}
l\lesssim l_{\rm crit}\sim {G\sigma\over \Omega^2} 
\sim \alpha^{-3}{\sigma\over\rho}
\ 
\end{equation}
collapse and virialize on the timescale
\begin{equation}
t_{\rm collapse}\sim \left({l\over G\sigma}\right)^{1/2}
 \ .
\label{eq:tcollapse}
\end{equation}
Perturbations with $l>l_{\rm crit}$ are stabilized by Keplerian shear
or, equivalently, by the Sun's tidal gravity.

Gravitational instabilities convert potential energy into kinetic energy of random motions.
The development of nonlinear overdensities requires that this energy dissipate at the 
collapse rate.  Otherwise the random velocity dispersion would be maintained near
the margin of stability
(Gammie 2001); i.e., $u\sim u_{\rm min}$.
 Inelastic collisions are the only
option for dissipating energy in a particle disk.  For the collision rate to match the collapse rate, the
particle disk would have to be optically thick , $\sigma/\rho s\gtrsim 1$.  An optically thick particle
disk might result from a collisional fragmentation cascade.

Only those clusters that originate from patches of size $l_*$, where
$l_*\ll l_{\rm crit}$, can contract
to form solid bodies
without getting rid of internal angular momentum. Here the asterisk subscript denotes quantities pertaining to the largest solid body that
can form without angular momentum loss. Its angular momentum per unit
mass, $(G\rho)^{1/2}s_*^2$, must be equal to the angular momentum per
unit mass of the patch from which it formed, $l_*^2\Omega$.
Combining this with mass conservation written as $\rho s_*^3\sim
\sigma l_*^2$, we arrive at
\begin{equation}
s_*\sim \alpha^{-3/2}{\sigma\over\rho}  \ 
\end{equation}
and
\begin{equation}
l_*\sim
\left({G\rho\over\Omega^2}\right)^{3/4}
{\sigma\over\rho}
\sim 
\alpha^{-9/4}{\sigma\over\rho} \sim \alpha^{3/4}l_{\rm crit} \, .
\end{equation}
Setting $\sigma$ to the value in the minimum mass solar nebula (Equation \ref{eq:mmsn}) gives
\begin{eqnarray}
s_*&\sim& 1 {\rm \  km}  \ 
\end{eqnarray} 
for the size of the bodies that form out of the disk,
independent of distance from the Sun.

Patches larger than $l_*$ have too much angular momentum
to collapse to a solid body.  Rather, they evaporate on the timescale
for two-body interactions.  In this context, we note that the escape
speed from bodies of size $s_*$,
\begin{equation}
(G\rho)^{1/2}s_*\sim \alpha^{-3} {\sigma\Omega\over \rho}
\sim {G\sigma\over\Omega}
 \ ,
\end{equation}
is that required to marginally stabilize the disk (Equation \ref{eq:uminfirst}).

The bodies of size $s_*$ formed by gravitational instability (described above)
are  ``first generation planetesimals" \cite{GW73}. This formation mechanism
was criticized by Weidenschilling (1980) and Weidenschilling \& Cuzzi (1993), who argued that gas prevents the settling
of the dust. Whether that statement is correct or not is still under debate \cite{YS02}.
However, this objection is less relevant for the formation of Uranus and Neptune, as
gas probably plays a minor role in the outer Solar System. The second stage
described by Goldreich \& Ward (1973)  does not materialize in the absence of gas drag.

\section{Appendix D: A More Complete Derivation of Dynamical Friction Cooling}
\label{sec:chandra}

Our derivation of Equations \ref{eq:chandra}--\ref{eq:uvs} is very
crude.  A more complete derivation of the dynamical friction drag
formula (Equation \ref{eq:chandra}) is obtained by integrating over both
the velocity distribution of the small bodies and all impact
parameters.  Aside from a logarithmic correction, this derivation
confirms Equation \ref{eq:chandra}.  Similar considerations apply to
Equations \ref{eq:vheat} and \ref{eq:uvs}.  We follow the
treatment of Binney \& Tremaine (1987, pp. 420--24).

When a small body passes a big body, it is deflected and thus
transfers momentum to the big body.  We calculate the momentum
transfer in the reference frame of the big body where we denote the
incoming velocity of the small body relative to the big body by ${\bld
U}_{\rm rel}$ and its impact parameter by $b$. In the limit
$GM/(bU_{\rm rel}^2)\ll 1$, the small body is deflected by an
angle
\begin{equation}
\theta\sim GM/(bU_{\rm rel}^2)\ll 1 \,\, . 
\end{equation}
The speed of the small body is unchanged by the deflection, but
its velocity vector is rotated by $\theta$, so its velocity is changed
in the direction transverse to its incoming velocity by
\begin{equation}
\Delta U_{\rm rel,\perp}/U_{\rm rel}\sim \theta \ 
\end{equation}
and decreased in the direction parallel to its incoming velocity by
\begin{equation}
\Delta U_{\rm rel,\parallel}/U_{\rm rel}\sim \theta^2 \ \ \Rightarrow
\Delta {\bld U}_{\rm rel, \parallel}\sim -\theta^2{\bld U}_{\rm rel} \
.
\end{equation}
(The subscript $\perp$ in this Appendix denotes a quantity transverse to
the incoming velocity; it should not be confused with the $\perp$ subscript
in Section \ref{subsec:inclinations}, where it denotes the vertical direction.)
Although $\Delta U_{\rm rel,\perp}>\Delta U_{\rm rel,\parallel}$,
$\Delta U_{\rm rel,\perp}$ does not contribute to cooling by dynamical
friction because it averages to nearly zero after many interactions.
Instead, it contributes to heating.  Therefore, for the
purpose of calculating dynamical friction drag, an interaction with a
small body changes the big body's momentum by
\begin{equation}
M\Delta {\bld v} =-m\Delta{\bld U}_{\rm rel,\parallel} \sim
\left(GM\over b U_{\rm rel}^2\right)^2 m{\bld U}_{\rm rel}\,\, .
\end{equation}
To determine how ${\bld v}$ changes over time, we integrate over
the flux of small bodies:
\begin{equation}
M{d{\bld v} \over dt} \sim \int bdb \int d^3{\bld U}f({\bld U})U_{\rm
rel} \left(GM\over b U_{\rm rel}^2\right)^2m{\bld U}_{\rm rel}  \,\, ,
\label{eq:df1}
\end{equation}
where $f({\bld U})$ is the distribution function of the 
small bodies and ${\bld U}_{\rm rel}\equiv {\bld U}-{\bld v}$

Integrating over $b$, we have $\int db/b\sim \ln(b_{\rm max}/b_{\rm
min}) \equiv \ln\Lambda$; equal parallel momentum transfer occurs
within each logarithmic decade of impact parameters.  The value of
$b_{\rm min}$ is that at which $\theta\sim 1$, and $b_{\rm max}$ is
given by the scale height of the disk of small bodies.  Typically,
$\ln\Lambda\sim 1$ in protoplanetary disks.  With this Coulomb
logarithm, Equation \ref{eq:df1} becomes
\begin{equation}
M{d{\bld v} \over dt} \sim
(GM)^2m\ln{\Lambda} \int d^3{\bld U}f({\bld U}) 
{{\bld U-\bld v}
\over \vert{\bld U-\bld v}\vert^3} \ .
\label{eq:dfx}
\end{equation}

The integral over $\bld U$ is, within a multiplicative constant,
equivalent to calculating the gravitational force on a test particle
at position $\bld{v}$ owing to an extended body whose mass density as a
function of position is $f({\bld U})$ \cite{BT87}.
 In a protoplanetary disk, $f({\bld U})$ may
be approximated as a triaxial Gaussian. The rms velocities in the
radial, azimuthal, and vertical dimensions are each comparable to
$u$, although they differ by order-unity factors.  The integral over
velocity yields $\sim -vf(0)$ in each dimension; this can be seen by
changing integration variables to $\bld w\equiv\bld U-\bld v$ and
expanding the Gaussian to lowest order in $v/u$.  If 
the integral were restricted to small bodies with $U \lesssim v$, then
 $(\bld U-\bld v)/\vert\bld U-\bld v\vert^3\sim -\bld v/v^3$. Thus
these low-velocity bodies contribute $\sim -vf(0)$ to the
integral, i.e., their fractional contribution to the integral is of
order unity, even though they represent only a small fraction,
$v^3/u^3$, of all the small bodies.  Indeed, when the
small-body distribution is isotropic, only bodies with $U<v$
contribute (Binney \& Tremaine 1987, Chandrasekhar 1943).  In our
case, however, the distribution is triaxial, and high-velocity bodies
are equally important.

Substituting $-vf(0)\sim -vn_s/u^3$ into Equation \ref{eq:dfx}, we find
\begin{equation}
{1\over v}{dv\over dt}\sim -G^2 Mm{n_s\over u^3}\ln\Lambda \sim
-\Omega{\sigma\over\rho R} \left({\vesc\over u}\right)^4 \ln
\Lambda\,\, ,
\label{eq:chandrareal}
\end{equation}
which, aside from the Coulomb logarithm, is Equation \ref{eq:chandra}.
We neglect this logarithm in our formulae, as it is of order
unity.  Equation \ref{eq:chandrareal} is often referred to as
Chandrasekhar's (1943) dynamical friction formula.

In our earlier crude derivation of Equation
\ref{eq:chandra}, we made two implicit assumptions that are not
quite correct.  First, we assumed that the dominant impact parameter
is $\sim GM/u^2$. In truth, all impact parameters within each
logarithmic decade, from $GM/u^2$ to the scale height of the disk,
contribute equally; this leads to the Coulomb logarithm.  Second, we
assumed that the dominant small bodies are those whose speed is $\sim
u$.  In fact, small bodies with speeds $\lesssim v$ make a fractional
contribution of order unity.

\begin{table} \begin{tabular}{ll}\hline
Quantity & Symbol \\ \hline\hline
Material density (of the Sun and of orbiting bodies) 
                          &$\rho\sim$ 1 \gm\cm$^{-3}$ \\ \hline
Solar radius              & $R_\odot$    \\
Solar mass                & $M_\odot\sim\rho R_\odot^3$    \\
Semimajor axis around the Sun & $a$    \\
Angular size of the Sun       & $\alpha\sim R_\odot/a$  \\
Orbital frequency around the Sun & $\Omega = (GM_\odot/a^3)^{1/2}$ \\
\hline
Big bodies' surface mass density &$\Sigma$ \\
Big bodies' volumetric number density &$n_b$\\
Big bodies' velocity dispersion &$v$ \\
Big bodies' radius  &$R$ \\
Big bodies' mass    &$M\sim\rho R^3$ \\
Big bodies' escape speed &$\vesc\sim (GM/R)^{1/2}\sim (G\rho)^{1/2}R$ \\
Big bodies' Hill radius         &$R_H\sim 
                     a(M/M_\odot)^{1/3} \sim R/\alpha$ \\ 
Big bodies' Hill velocity  &$v_H \sim \vesc\alpha^{1/2}
                                 \sim \Omega R_H $ \\
\hline
Small bodies' surface mass density &$\sigma$ \\
Small bodies' volumetric number density &$n_s$\\
Small bodies' velocity dispersion &$u$ \\
Small bodies' radius  &$s$ \\
Small bodies' mass    &$m\sim\rho s^3$ \\
Small bodies' escape speed &$u_{\rm esc}\sim (G\rho)^{1/2}s$ \\
\hline
Radius of bodies that condense out of a cold disk
& $s_*\sim (\sigma/\rho)\alpha^{-3/2}$ \\
\hline
\end{tabular}\\[0.5ex] 
\caption{Main symbols used in text}
\label{table:symbols}
\end{table}

\section{Acknowledgments}

We thank M. Brown, E. Chiang, L. Dones, M. Duncan, J. Goodman, 
S. Ida,
S. Kenyon, 
E. Kokubo,
J. Makino, N. Murray, R. Rafikov,  D. Stevenson, 
E. Thommes, S. Tremaine, 
and A. Youdin
for useful discussions. This research was supported in part
by NSF grants AST-0098301 and PHY-0070928 and by NASA grant NAG5-12037.

\end{document}